\documentclass[10pt,aps,prc,floatfix,twocolumn,superscriptaddress,nofootinbib]{revtex4-2}

\usepackage[table,dvipsnames]{xcolor}
\usepackage{amsfonts,amsmath,amssymb,bm}
\usepackage{graphicx}
\usepackage[utf8]{inputenc}
\usepackage{xspace}
\usepackage{dcolumn}
\usepackage{multirow}
\newcolumntype{d}[1]{D{.}{.}{#1}}
\usepackage{cellspace}
\usepackage{isotope}
\usepackage{xparse}
\usepackage{physics}
\usepackage{dsfont}
\usepackage[pdfpagelabels, pdfencoding=auto, psdextra]{hyperref}
\hypersetup{%
 pdfsubject=Paper,
 pdfkeywords={nuclear physics} 
 unicode = true,
 breaklinks = true,
 colorlinks = true,
 linkcolor = blue,
 menucolor = blue,
 citecolor = blue,
 urlcolor = blue
}
\usepackage{enumitem}
\usepackage{mathtools}

\usepackage{tikz}
\usepackage{bm}
\usetikzlibrary{fit}
\usetikzlibrary{calc,trees,positioning,arrows,chains,shapes.geometric,%
    decorations.pathreplacing,decorations.pathmorphing,shapes,%
    matrix,shapes.symbols}

\tikzset{
  >=stealth',
  punktchain/.style={
    rectangle, 
    rounded corners, 
    draw=black, very thick,
    text width = 0.95 \columnwidth, 
    minimum height=1em, 
    text centered, 
    on chain},
    line/.style={draw, thick, <-},
    element/.style={
    tape,
    top color=white,
    bottom color=blue!50!black!60!,
    minimum width=8em,
    draw=blue!40!black!90, very thick,
    text width=10em, 
    minimum height=3.5em, 
    text centered, 
    on chain},
  every join/.style={->, thick,shorten >=0pt},
  decoration={brace},opacity=0.5,thick,inner sep=0pt,
  tuborg/.style={decorate},
  tubnode/.style={midway, right=2pt},
  highlight/.style={rectangle,rounded corners,fill=red!15,draw,
    fill opacity=0.5,thick,inner sep=0pt}
}

\graphicspath{{./figures/}}

\newcommand{\orcid}[1]{\href{https://orcid.org/#1}{\includegraphics[scale=0.055]{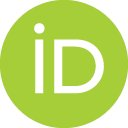}}}

\newcommand{\paramVec}{\vb*{\theta}} 
\newcommand{\fmisq}{\, \text{fm}^{-2}}
\newcommand{\fmsq}{\, \text{fm}^{2}}

\newcommand{\ntheta}{n_{\paramVec}}
\newcommand{\Amat}{\vb{A}}  
\newcommand{\tAmat}{\widetilde\Amat}
\newcommand{\svec}{\vb{s}}   
\newcommand{\cvec}{\vb{c}}   
\newcommand{\yvec}{\vb{y}}   
\newcommand{\tsvec}{\widetilde\svec}

\newcommand{\Mmat}{\vb{M}}  
\newcommand{\Dmat}{\vb{D}}  
\newcommand{\Umat}{\vb{U}}  
\newcommand{\Vmat}{\vb{V}}  
\newcommand{\Xmat}{\vb{X}}  
\newcommand{\Ymat}{\vb{Y}}  

\newcommand{\residual}{\vb{r}}
\newcommand{\ytilde}{\tilde{\vb{y}}}

\newcommand{\ATensor}{\vb{A}^{(\theta)}}

\newcommand{\sTensor}{\vb{S}^{(\theta)}}

\newcommand{\exactErrorVec}{\vb{e}} 
\newcommand{\romResidual}{\vb{r}}
\newcommand{\romResidualY}{\rvec_{\vb{Y}}}
\newcommand{\rvec}{\vb{r}}   
\newcommand{\KnXi}[2]{\mathcal{G}_{#1}^{#2}}

\newcommand{\oneSzero}{${}^1\text{S}_0$\xspace}  %
\newcommand{\threePzero}{${}^3\text{P}_0$\xspace}  %
\newcommand{\fm}{\, \text{fm}}
\newcommand{\MeV}{\, \text{MeV}}

\newcommand{\nTwoLO}{N$^2$LO\xspace}
\newcommand{\chiEFT}{$\chi$EFT\xspace}

\begin{document}

\title{Greedy Emulators for Nuclear Two-Body Scattering}

\author{J.~M. Maldonado~\orcid{0009-0009-8872-2385}}
\email{jm998521@ohio.edu}
\affiliation{Department of Physics and Astronomy, \href{https://ror.org/01jr3y717}{Ohio University}, Athens, Ohio~45701, USA}

\author{C.~Drischler~\orcid{0000-0003-1534-6285}}
\email{drischler@ohio.edu}
\affiliation{Department of Physics and Astronomy, \href{https://ror.org/01jr3y717}{Ohio University}, Athens, Ohio~45701, USA}
\affiliation{\href{https://ror.org/03r4g9w46}{Facility for Rare Isotope Beams}, \href{https://ror.org/05hs6h993}{Michigan State University}, East Lansing, Michigan~48824, USA}

\author{R.~J. Furnstahl~\orcid{0000-0002-3483-333X}}
\email{furnstahl.1@osu.edu}
\affiliation{Department of Physics, \href{https://ror.org/00rs6vg23}{The Ohio State University}, Columbus, Ohio 43210, USA}

\author{P.~Mlinari{\'c}~\orcid{0000-0002-9437-7698}}
\email{mlinaric@vt.edu}
\affiliation{Department of Mathematics, \href{https://ror.org/02smfhw86}{Virginia Tech}, Blacksburg, Virginia 24061, USA}

\date{\today}

\begin{abstract}  

Applications of reduced basis method emulators are increasing in low-energy nuclear physics because they enable fast and accurate sampling of high-fidelity calculations, enabling robust uncertainty quantification. 
In this paper, we develop, implement, and test two model-driven emulators based on the (Petrov-)Galerkin projection using the prototypical test case of two-body scattering with the Minnesota potential and a more realistic local chiral potential. 
The high-fidelity scattering equations are solved with the matrix Numerov method, a reformulation of the popular Numerov recurrence relation for solving special second-order differential equations as a linear system of coupled equations.
A novel error estimator based on reduced-space residuals is applied to an active learning approach (a greedy algorithm) to choosing training samples (``snapshots'') for the emulator and contrasted with a proper orthogonal decomposition (POD) approach.
Both approaches allow for computationally efficient offline-online decompositions, but the greedy approach requires much fewer snapshot calculations.
These developments set the groundwork for emulating scattering observables based on chiral nucleon-nucleon and three-nucleon interactions and optical models, where computational speed-ups are necessary for Bayesian uncertainty quantification.
Our emulators and error estimators are widely applicable to linear systems.

\end{abstract}

\maketitle

\section{Introduction}
\label{sec:intro}

With the development of precise and accurate computational tools to confront new experimental results or robustly extrapolate to as-yet unmeasured domains, uncertainty quantification (UQ) has become a requirement in low-energy nuclear theory.
This UQ typically involves Bayesian statistical methods for calibration, sensitivity analyses, error propagation, and more.
These methods require many samples from high-fidelity, or full-order model (FOM), calculations with different Hamiltonian parameters, energies, or other control parameters, which may become prohibitively expensive or at least detrimental to efficient analyses.
Fast \& accurate emulators, or reduced-order models (ROMs), can solve this problem.
An effective approach is to use reduced basis method (RBM) emulators, which are trained using a selection of high-fidelity solutions, generically called snapshots, with parameters chosen such that a subspace is spanned that accurately represents the relevant part of the full solution space (see Refs.~\cite{Melendez:2022kid,Drischler:2022ipa,Duguet:2023wuh} for details and visualizations of the RBM approach).
But how should one choose the snapshots?

\begin{figure}[b!]
    \centering
    \includegraphics[width=\columnwidth]{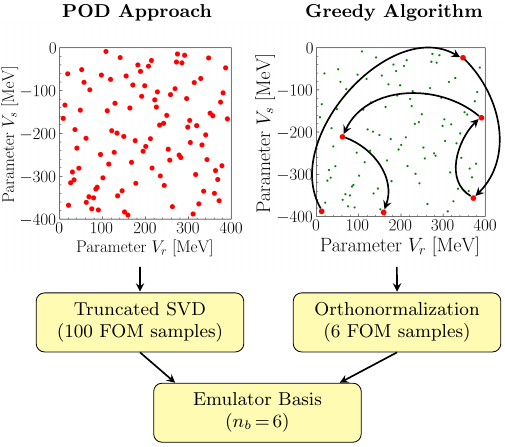}
    \caption{%
    Illustrative comparison between the POD approach and the greedy method of constructing an emulator basis.
    The POD approach uses snapshots from many parameter values (red points) and a truncated SVD to produce a reduced orthonormal basis (here, the basis size is $n_b = 6$).
    The greedy method uses far fewer snapshot solutions by iteratively adding to the emulator basis only the FOM solution at the parameter value (following the arrows) that maximizes the estimated error across the training set (green points).
    This error estimation does not require expensive, high-fidelity calculations at the training points.
    These snapshots are then orthonormalized, e.g., by the Gram-Schmidt process.
    }
    \label{fig:pod-vs-greedy}
\end{figure}

Two strategies for selecting snapshots are common in the reduction of parametric problems~\cite{hesthaven2015certified,Quarteroni:218966,Benner20201}:
\begin{enumerate}
    \item Sample the parameter space with a space-filling algorithm, such as Latin hypercube sampling (LHS), and then compress the resulting basis with a (truncated) singular value decomposition (SVD). This is known as proper orthogonal decomposition (POD). 
    \item Choose snapshots iteratively by minimizing the immediate global error across the parameter space; this is known as a greedy algorithm. This approach requires a robust error estimator~\cite{hesthaven2015certified},
    which is also needed for a complete UQ error budget.
\end{enumerate}
Contrasting visualizations of these approaches are shown in Fig.~\ref{fig:pod-vs-greedy}. 
Both approaches have been implemented for quantum systems, but there is still much to explore.%
\footnote{%
Note also that these two approaches can be enriched by additional information on the important regions in the parameter space that, e.g., may come from prior knowledge or Bayesian importance weights~\cite{Jiang:2022oba}.}
In this work, we develop and test a multidimensional greedy algorithm for two-body scattering that uses a novel error estimator.

Previous applications of the POD approach in nuclear physics have been made in Refs.~\cite{Giuliani:2022yna,Odell:2023cun} with the publicly available software package \texttt{ROSE} from the BAND collaboration~\cite{BAND_Framework}.  
Although these efforts have shown great success, a potential drawback is that many FOM calculations are required for training the emulator.
This is not always feasible, for example, for coupled cluster calculations of nuclear structure~\cite{PhysRevLett.123.252501} or for emulating few-body scattering~\cite{Zhang:2021jmi}. 
In such situations, keeping the number of high-fidelity solutions to a minimum is important. 

The active learning approach using a greedy algorithm generally minimizes the number of snapshots needed by iteratively adding snapshots until a prescribed accuracy requirement is met.
Previous work on greedy algorithms in quantum physics was done by Sarkar and Lee~\cite{Sarkar:2021fpz} and Bonilla \textit{et al.}~\cite{Bonilla:2022rph}. 
These works demonstrate that a greedy algorithm is suitable for constructing a basis for a one-dimensional parameter space for multiple types of emulators.
A recent application of a greedy algorithm to quantum spin systems that yields an efficient mapping of ground-state phase diagrams was made in Refs.~\cite{Herbst2022,Brehmer2023}.

The present work develops, implements, and tests two projection-based model-driven emulators, a Galerkin Reduced-Order Model (G-ROM) emulator and a least-squares Petrov-Galerkin ROM (LSPG-ROM) emulator. These two emulators are used in conjunction with a novel error estimator based on reduced space residuals.
Both the emulators and their error estimators use an efficient offline-online decomposition to allow for computational speed-ups.
The FOM is implemented using the matrix Numerov method.
This implementation sets the stage for future work in rigorous UQ for two- and three-body nuclear scattering.
We start with the controlled case of a simple Minnesota potential~\cite{THOMPSON197753} 
to illustrate the process. 
Subsequently, we generalize to more realistic two-body potentials derived from chiral effective field theory (\chiEFT) to test and compare the resulting emulators against alternative approaches.

\begin{table}[tb]
\renewcommand{\arraystretch}{1.2}
\setlength{\tabcolsep}{7pt}
\caption{
Acronyms used in this work.
}
\label{tab:acronym}
\begin{ruledtabular}
\begin{tabular}{l>{\raggedright\arraybackslash}p{5.9cm}} 
Acronym & Stands for \\ 
\colrule 
\chiEFT & Chiral effective field theory\\
EIM & Empirical interpolation method [Sec.~\ref{subsec:rse}]\\
FOM & Full-order model [Sec.~\ref{sec:fom}]\\
G-ROM & Galerkin ROM [Sec.~\ref{sec:G-ROM}]\\
GT+ & Gezerlis, Tews \textit{et al.}~\cite{Gezerlis:2014zia}\\
LEC & Low-energy constant\\
LHS & Latin hypercube sampling~\cite{Iman:2008}\\
LSPG-ROM & Least-squares Petrov-Galerkin ROM [Sec.~\ref{sec:lspg}]\\
\nTwoLO & Next-to-next-to-leading order \\
ODE & Ordinary differential equation [Sec.~\ref{sec:fom}]\\
POD & Proper orthogonal decomposition [Sec.~\ref{sec:pod}]\\
RBM & Reduced basis method 
to construct ROMs~\cite{hesthaven2015certified,Quarteroni:218966,Benner2020Volume1DataDriven,Benner20201,Benner2020Volume3Applications}\\
ROM & Reduced-order model  [Sec.~\ref{sec:rom}]\\
RSE & Radial Schr{\"o}dinger equation [Eq.~\eqref{eq:rse}]\\ 
SCM & Successive constraint method [Sec.~\ref{sec:error_estimates}] \\
SVD & Singular value decomposition [Sec.~\ref{sec:pod}] \\
UQ & Uncertainty quantification 
\end{tabular}
\end{ruledtabular}
\end{table}

\begin{table}[tb]
\renewcommand{\arraystretch}{1.2}
\setlength{\tabcolsep}{6pt}
\caption{
Notation used in this work.
}
\label{tab:notation}
\begin{ruledtabular}
\begin{tabular}{lp{6.0cm}} 
Notation & Description \\ 
\colrule 
$n_b$ & Number of reduced basis elements [Eq.~\eqref{eq:rom_reduction} and Fig.~\ref{fig:pod-vs-greedy}] \\
$\paramVec$, $n_\theta$ & Contains the $\ntheta$ low-energy couplings or other model parameters as its components  \\
$a,b$ & Affine decomposition indices $\in [0,1,\dots,n_\theta]$ \\
$i, j$ & FOM indices $\in [1,2,\dots,N-1]$ \\
$u,v,w$ & ROM indices $u,v\in [1,2,\dots,n_b]$ and $w \in [1,2,\ldots,n_{\Ymat}]$\\
$n$, $N$ & Grid index $\in[0,1,2,\ldots,N]$, with the number of grid points $N+1$ \\
$\Xmat$ & Snapshot matrix ($N \times n_b$) having the emulator basis vectors as its columns \\
$\Ymat$ & Projection matrix of the LSPG-ROM [Eq.~\eqref{eq:def_Ymat}], with dimensions $N \times n_{\Ymat}$\\
$\delta_\ell$ & Scattering phase shift at energy $E$, not to be confused with the  Kronecker delta $\delta_{i,j}$ \\
$r$, $r_n$, $h$ & Radial coordinate $r$ sampled at equidistant grid points $r_n$ with step size $h$ \\ 
$\ell$ & Relative orbital angular momentum quantum number \\
$\mu$, $p$, $E$ & Reduced mass, relative momentum $p = \sqrt{2\mu E}$ associated with the energy $E > 0$\\
$\phi_\ell(r; \paramVec)$ & Radial wave function, solution to homogeneous RSE [Eq.~\eqref{eq:rse}]\\
$\chi_\ell(r; \paramVec)$  & Scattered wave function [Eq.~\eqref{eq:def_chi}]\\
$K_{\ell}(\paramVec) $ & Scattering $K$ matrix element [Eq.~\eqref{eq:Kmat_delta}] \\
$F_{\ell}, G_{\ell}$ & Riccati-Bessel functions; i.e., independent free-space solutions of homogeneous RSE\\
$g(r;\paramVec)$ & Homogeneous part of the RSE [Eq.~\eqref{eq:def_fry}]\\
$s(r;\paramVec)$ & Inhomogeneous part of the RSE [Eq.~\eqref{eq:def_fry}]\\
$ \Amat(\paramVec)$ & FOM (i.e., Numerov) matrix [Eq.~\eqref{eq:linear_system}] \\
$\vb{y}(\paramVec) $ & FOM solution $\{ y_2, y_3,\ldots, y_N\}$ [Eq.~\eqref{eq:linear_system}] \\
$\vb{s}(\paramVec) $ & FOM right-hand side vector [Eq.~\eqref{eq:linear_system}] \\
$\cvec(\paramVec)$ & To-be-determined ROM coefficient vector\\
$\exactErrorVec(\paramVec)$ & emulator error (to be estimated) [Eq.~\eqref{eq:rom_error}]\\
$\romResidual(\paramVec)$ & Emulator residual [Eq.~\eqref{eq:residual}]   \\
$\romResidualY(\paramVec)$ & Emulator residual projected onto the subspace of residuals [Eq.~\eqref{eq:rom_residual_Y}] \\
$\zeta$ & Indicator variable: $\zeta = 0$ ($\zeta = 1$) for the homogeneous (inhomogeneous) RSE\\
$r_{m_t}$, $\tau$ & Grid points with indices $\{m_1, m_2, \ldots, m_\tau\}$ used to extract the $K$ matrix element
\end{tabular}
\end{ruledtabular}
\end{table}

The remainder of this paper is organized as follows:
In Sec.~\ref{sec:fom}, we describe and benchmark the FOM with the matrix Numerov method for two-body scattering.
In accord with the general philosophy of this work as providing a prototype for emulation, we give sufficient detail to enable both reproduction of our results and extension to other problems.
The ROMs and error estimators are presented in Sec.~\ref{sec:rom}, with explicit descriptions of the offline-online divisions.
Snapshot selection via the POD or greedy algorithm is discussed in Sec.~\ref{sec:snapshot_selection}.
Section~\ref{sec:results} discusses our results, and Sec.~\ref{sec:summary_outlook} finishes this article with a summary and outlook.
Several appendices provide additional information: 
Appendix~\ref{app:init_gonzalez} discusses applications of the Numerov method to initial value problems; 
Appendix~\ref{app:other_emulators} introduces the ``all-at-once Numerov method,'' a useful method for emulating the scattering $T$-matrix element directly, if one can tolerate unitary violation (and restoration) of the scattering $S$ matrix, and similar variations of the presented emulators; and
Appendix~\ref{app:prestore_scalar_residual} details how the scalar emulator error can be efficiently prestored in the offline stage.
We use natural units in which $\hbar = c = 1$ and use upper and lower case letters to denote matrices and vectors, respectively, both typeset in boldface. 
The main acronyms used in this manuscript are summarized in Table~\ref{tab:acronym} and the notation in Table~\ref{tab:notation}.  
Source codes for reproducing and extending our results are publicly available on GitHub~\cite{BUQEYEsoftware}.

\section{Full-Order Model: High-Fidelity ODE solver}
\label{sec:fom}

In this section, we introduce the partial-wave decomposed radial Schr{\"o}dinger equation (RSE) for two-body scattering and describe how to solve this ordinary differential equation (ODE) using the matrix Numerov method, a reformulation of the popular Numerov recurrence relation as a linear system of coupled equations.
This method is our FOM. 
We explain this method in detail because of its expected usefulness for extended problems (such as three-body scattering).

\subsection{Radial Schr{\"o}dinger equation}
\label{subsec:rse}

We consider two-body scattering in coordinate space with short-range interactions.\footnote{The extension to the long-range Coulomb interaction is straightforward; see, e.g., Refs.~\cite{Furnstahl:2020abp,Melendez:2021lyq}.} 
While our framework is generally applicable, e.g., to nonlocal and complex-valued (optical) potentials, we focus in this work on neutron-proton ($np$) scattering with local potentials as a proof of principle.
In a plane-wave partial wave basis without channel coupling, the radial wave function $\phi_\ell(r)$ is given by the (regular) solutions of the linear \emph{homogeneous} RSE:
\begin{equation} \label{eq:rse}
    \dv[2]{r} \phi_\ell(r; \paramVec) = \left[2\mu V_\ell(r; \paramVec) + \frac{\ell (\ell + 1)}{r^2} - p^2 \right]  \phi_\ell(r; \paramVec)  \,,
\end{equation}
where $r$ is the radial coordinate, $\mu$ the reduced mass, and $p = \sqrt{2\mu E}$ the relative momentum of the scattering particles associated with the center-of-mass energy $E > 0$, and $\ell \geqslant 0$ the quantum number for the relative orbital angular momentum.
The coordinate-space potential $V_\ell(r; \paramVec)$ depends on the parameter vector $\paramVec$, which is typically calibrated to experimental data and should not be confused with the scattering angle.
For example, $\paramVec$ may contain the low-energy constants (LECs) of a chiral potential or the parameters of an optical model.

To construct a computationally efficient offline-online decomposition for the emulator, where computationally intensive operations are performed once upfront in the offline stage and then reused in the online stage, we demand that $V_\ell(r; \paramVec)$ has an affine parameter dependence,
\begin{equation} \label{eq:affine_potential}
    V_\ell(r; \paramVec) = \sum_{a=0}^{\ntheta} h_a^{(\ell)}(\paramVec) \mathcal{V}_a^{(\ell)}(r) \,,
\end{equation}
where the $h_a^{(\ell)}(\paramVec)$ are parameter-dependent functions while the $\mathcal{V}_a^{(\ell)}(r)$ are parameter-independent.
Although the functions $h_a^{(\ell)}(\paramVec)$ are only required to be smooth, not necessarily linear in $\paramVec$, we focus here on $h_a^{(\ell)}(\paramVec) = \paramVec_a$, which applies to the short-range nucleon-nucleon (NN) contact interactions of chiral interactions. 
Note that we introduce $\paramVec_0 \equiv 1$ to accommodate a $\paramVec$-independent (i.e., constant) term in the affine decompositions. We will not emulate solutions in this auxiliary dimension.
If $V_\ell(r; \paramVec)$ does not have an affine parameter dependence, one can use the empirical interpolation method (EIM) to approximately cast $V_\ell(r; \paramVec)$ into the form of Eq.~\eqref{eq:affine_potential}, as demonstrated in Ref.~\cite{Odell:2023cun} for optical models.

At a given $E$, the solutions of the RSE~\eqref{eq:rse} are required to be regular, i.e., $\phi_\ell(r=0; \paramVec) = 0$, and normalized such that their asymptotic limit is parametrized by\footnote{If $\phi_\ell(r; \paramVec)$ is a solution to the (linear) homogeneous RSE~\eqref{eq:rse} then $C\phi_\ell(r; \paramVec)$ with an arbitrary constant $C$ is also a solution.}
\begin{equation} \label{eq:asymLimit}
    \phi_{\ell}(r;\paramVec) \sim  \frac{1}{p} \Bigl[ F_\ell(pr) + K_{\ell}(\paramVec) \, G_\ell(pr) \Bigr] \quad (r \to \infty)\,, 
\end{equation}
where $F_\ell(z) = z j_\ell(z)$ and $G_\ell(z) = -z \eta_\ell(z)$ are the Riccati-Bessel functions.
They are the two independent free-space solutions, proportional to the spherical Bessel functions and von Neumann functions,  $j_\ell(z)$ and $\eta_\ell(z)$, respectively.
We choose to parametreize the asymptotic limit~\eqref{eq:asymLimit} in terms of the $K_{\ell}$ matrix element, which is related to the scattering phase shift via\footnote{For brevity, we omit the momentum dependence of the $K_{\ell}$ (and $S_{\ell}$) matrix element and associated phase shift.} 
\begin{equation} \label{eq:Kmat_delta}
   K_{\ell}(\paramVec) = \tan \delta_{\ell}(\paramVec) \,. 
\end{equation}
For NN potentials, this parametrization guarantees that the (FOM and ROM) calculations are real-valued and that the ROM approximations preserve the unitarity of the partial-wave scattering matrix element $S_{\ell}$ (see Sec.~\ref{sec:rom} for more details).
But other choices, such as the parameterization in terms of $S_{\ell}$, are equally valid and can be straightforwardly obtained from the $K_{\ell}$ matrix element parametrization~\eqref{eq:asymLimit} via linear fractional transformations, as discussed in detail in Ref.~\cite{Drischler:2021qoy}.

To make the boundary condition~\eqref{eq:asymLimit} explicit, we write
\begin{equation} \label{eq:def_chi}
    \phi_\ell(r; \paramVec) = \frac{1}{p} [ F_\ell(pr) + \chi_\ell(r; \paramVec)] \,,
\end{equation}
and express the RSE in terms of the scattered wave function, $\chi_\ell(r;\paramVec)$, resulting in the \emph{inhomogeneous} RSE
\begin{equation} \label{eq:inhom_rse}
\begin{split}
    \dv[2]{r} \chi_\ell(r; \paramVec) 
    &=  
    \left[
        2 \mu V_\ell(r; \paramVec) + \frac{\ell (\ell + 1)}{r^2}  - p^2
    \right] \chi_\ell(r; \paramVec) \\ 
    & \quad + 2 \mu V_\ell(r; \paramVec) \, F_\ell(pr) \,,
    \end{split}
\end{equation}
subject to the initial condition $\chi_\ell(r=0; \paramVec) = 0$ and, combined with Eq.~\eqref{eq:def_chi}, the boundary condition~\eqref{eq:asymLimit}.
In the absence of a potential, the solution to the inhomogeneous RSE~\eqref{eq:inhom_rse} vanishes by construction.

Another motivation for considering the inhomogeneous RSE
is its better numerical behavior (explained below).
Using the Frobenius method,\footnote{For a comprehensive discussion of the Frobenius method applied to second-order ODEs, see Sec.~4.2 in Ref.~\cite{teschlordinary}. The special case relevant to this work is obtained by setting $p(z) = 0$ in Eq.~(4.20).}
assuming that $V_{\ell}(r; \paramVec)$ is bounded around $r = 0$,
we find that the regular solution of the homogeneous RSE has the form
\begin{equation}
    \phi_{\ell}(r; \paramVec)
    = C r^{\ell + 1}
    + \frac{2 \mu V_{\ell}(0; \paramVec) - p^2}{2 (2 \ell + 3)} C r^{\ell + 3}
    + O(r^{\ell + 4}),
\end{equation}
for some constant $C$.
In particular, the dominant term around $r = 0$ is of the order $r^{\ell + 1}$, while there is no term of order $r^{\ell + 2}$.
The same holds for $F_{\ell}(p r)$ around $r = 0$.
Therefore, for a particular value of $C$, the subtraction in $\chi_{\ell}(r; \paramVec) = p\, \phi_{\ell}(r; \paramVec) - F_{\ell}(p r)$ leads to $\chi_{\ell}(r; \paramVec)$ having the dominant term around $r = 0$ of the order $r^{\ell + 3}$ (instead of $r^{\ell + 1}$).

\subsection{Numerov ODE solver}
\label{subsec:fom}

The homogeneous and inhomogeneous RSEs discussed in the previous section are parametrized second-order ODEs that do not depend on the derivative of the solution function.
They can be cast into the special form:
\begin{equation} \label{eqn:formal-numerov}
    y''(r;\paramVec) = f(r, y(r;\paramVec); \paramVec) \,, 
\end{equation}
with the function $f$ being linear in $y$ 
\begin{equation}\label{eq:def_fry}
    f(r, y; \paramVec) = -g(r;\paramVec) y + s(r;\paramVec) \,,
\end{equation}
subject to the initial condition $y(r_0;\paramVec) = y_0$ and boundary condition~\eqref{eq:asymLimit}. 
For the inhomogeneous RSE~\eqref{eq:inhom_rse}, one identifies
\begin{subequations} \label{eqn:numerov-mapping}
\begin{align} 
    y(r;\paramVec) &=  \chi_\ell(r;\paramVec) \,,\\
    -g(r;\paramVec)  &=  2 \mu \, V_\ell(r; \paramVec) + \frac{\ell (\ell + 1)}{r^2} - p^2 \,, \label{eq:mapping_g(r)}\\
    s(r;\paramVec) &= \zeta \,2 \mu \, V_\ell(r; \paramVec) \, F_\ell(pr) \,,
\end{align}
\end{subequations}
and for the homogeneous RSE~\eqref{eq:rse}, one finds $y(r) = \phi_\ell(r)$, $g(r)$ as defined in Eq.~\eqref{eq:mapping_g(r)}, and $s(r) = 0$.
We introduce the indicator variable $\zeta$ to handle the inhomogeneous ($\zeta =1$) and homogeneous case ($\zeta =0$) simultaneously.
For simplicity, the subscript $\ell$ indicating the given partial-wave channel will be omitted from here on.

Many numerical methods exist to solve the special second-order ODE~\eqref{eqn:formal-numerov} accurately and efficiently, such as the Numerov and Runge-Kutta (RK) methods.  
Here, we consider the general class of ODE solvers that can be expressed as a linear system of coupled equations, excluding adaptive methods such as RK45.
We choose the Numerov method as our FOM solver for this proof of principle study because of its high accuracy at low computational complexity and popularity in nuclear physics.

\subsubsection{Numerov recurrence relation}
\label{sec:numerov_recurrence}

The Numerov method of solving the ODE~\eqref{eqn:formal-numerov} provides an explicit, multi-step integration rule that achieves fourth-order accuracy.\footnote{The standard Numerov method may be unstable for bound-state calculations~\cite{friar-numerov} as the wave function vanishes as $r\to \infty$. We have not observed these issues in scattering calculations.}
Assuming space discretized on the equidistant grid $\{r_n\}_{n=0}^{N}$ with step size $h = r_{n+1} - r_{n}$ for $n = 0, 1,\ldots, N-1$, the Numerov method solves the ODE~\eqref{eqn:formal-numerov} by constructing a sequence $\{y_n\}_{n=0}^{N}$, with $y_n \approx y(r_n)$, using the recurrence relation (e.g., see also Refs.~\cite{Numerov19272301903,Noumerov84.8.592,BLATT1967382,SIMOS1991201})
\begin{equation} \label{eq:numerov_recurrence}
    \begin{split}
        \KnXi{n+1}{(1)} \, y_{n+1} &= 
        2 \KnXi{n}{(-5)} \, y_n
        - \KnXi{n-1}{(1)} \, y_{n-1}\\
        & \quad + \frac{h^2}{12} \left( s_{n+1} + 10 s_n + s_{n-1} \right) \,,
    \end{split}
\end{equation}
given initial values for $y_0$ and $y_1$. 
Here, we use the short-hand notations $g_n \equiv g(r_n)$ and $s_n \equiv s(r_n)$ for these functions evaluated on the grid, and 
\begin{equation}
    \KnXi{n}{(\xi)} = 1 + \xi \, \frac{h^2}{12} \, g_n \,.
\end{equation}
How to initialize the recurrence relation~\eqref{eq:numerov_recurrence} when
solving the RSE?%
\footnote{%
    For completeness, we discuss in Appendix~\ref{app:init_gonzalez} an approach for determining $y_1$ given initial values for $(y_0,y'_0)$ using a (truncated) series expansion of $y(r)$ about $r_0$. 
    The series coefficients are obtained by finite differencing.
}
Here, we use the grid points $r_n = n h$, with $n = 0, 1, \ldots, N$,
so $y_0 = 0$.
When solving the (linear) homogeneous RSE,
we can arbitrarily choose $y_1 = 1$ as it will be rescaled after applying the
Numerov recurrence relation by imposing the asymptotic limit
parametrization~\eqref{eq:asymLimit}.
For solving the inhomogeneous RSE, we set $y_0 = y_1 = 0$,
which introduces an error of order $O(h^{\ell + 3})$,
based on the analysis in Sec.~\ref{subsec:rse} using the Frobenius method.
At this point, the sequence $\{y_n\}_{n=0}^{N}$ still has to be matched to the asymptotic limit parametrization~\eqref{eq:asymLimit}, as described next.

Another question arises in evaluating $\KnXi{0}{(1)} \, y_0$
appearing in Eq.~\eqref{eq:numerov_recurrence} for $n = 1$ and $\ell > 0$,
since in that case, $g_0$ is not defined.
We interpret $\KnXi{0}{(1)} \, y_{0}$ as the limit
\begin{equation}
	\lim_{r \to 0^+}
    \left( 1 + \frac{h^2}{12} \, g(r) \right)
    y(r) \,.
\end{equation}
Using the results of the Frobenius method and some algebraic manipulation, for the homogeneous RSE,
we find that the limit is $-\frac{y_1 + O(h^3)}{6} \delta_{\ell,1}$,
which we approximate as $-\frac{y_1}{6} \delta_{\ell,1}$.
For the inhomogeneous RSE, we similarly find that the limit is zero for all $\ell \geqslant 0$,
which coincides with $-\frac{y_1}{6} \delta_{\ell,1}$ since we set $y_1 = 0$.

Once the recurrence relation is solved, one must still impose the boundary condition~\eqref{eq:asymLimit} regardless of whether the homogeneous or inhomogeneous RSE is considered.
To this end, we obtain $\phi(r_{m_\tau})$ at $\tau\geqslant 2$ different matching radii $\{r_{m_t}\}_{i=1}^\tau$, where the indices $m_t$ are chosen such that $r_{m_t}$ is located outside the range of the potential, by solving first the RSE and then the least-squares problem:
\begin{equation} \label{eq:lsq_matching}
\frac{1}{p} 
    \begin{bmatrix}
        F(p r_{m_{1}})  & G(p r_{m_{1}}) \\
        F(p r_{m_{2}})  & G(p r_{m_{2}}) \\
        \vdots & \vdots \\
        F(p r_{m_{\tau}})  & G(p r_{m_{\tau}}) 
    \end{bmatrix}
    \begin{bmatrix} a(\paramVec) - \zeta p \\ b(\paramVec) \end{bmatrix} =
        \begin{bmatrix}
        y_{m_{1}} \\
        y_{m_{2}}  \\
        \vdots  \\
       y_{m_{\tau}} 
    \end{bmatrix}\,,
\end{equation}
where $\zeta = 0$ ($\zeta = 1$) when solving the homogeneous (inhomogeneous) RSE.
With the least-squares solution, one can then determine 
\begin{subequations} \label{eq:K_delta_a_b}
\begin{align}
K(\paramVec) &= \frac{b(\paramVec)}{a(\paramVec)} \,,\\
\delta(\paramVec) &= \operatorname{atan2}\left[b(\paramVec), a(\paramVec) \right] \,. \label{eq:delta_from_K_a_b}
\end{align}
\end{subequations}
%
To obey the imposed boundary condition, the obtained (unmatched) Numerov solution has to be transformed as
\begin{equation}
y_n^{(\text{matched})} = 
\frac{y_n + \zeta \left\{ (p - 1)y_n - [a(\paramVec) - p] F(pr_n) \right\} }{a(\paramVec)} \,.
\end{equation}
%

A common alternative approach to the described matching procedure is to compute the inverse logarithmic derivative with respect to $r$,
\begin{equation}
    R_{m_t} = \frac{y_{m_t} + \zeta F(pr_{m_t})}{y'_{m_t} + \zeta F'(pr_{m_t})} \,,
\end{equation}
for all $m_t$, allowing one to determine
\begin{align}
    K(\paramVec; r_{m_t}) &= -\frac{F(pr_{m_t}) - R_{m_t} F'(pr_{m_t}) }{G(pr_{m_t}) - R_{m_t} G'(pr_{m_t})} \,,\\
    a(\paramVec; r_{m_t}) &= p \frac{y_{m_t} + \zeta F(pr_{m_t})}{F(pr_{m_t}) + K(\paramVec; r_{m_t}) G(pr_{m_t})} \,.
\end{align}
The $K(\paramVec)$ matrix element and scaling factor $a(\paramVec)$ can then be extracted by averaging the results; e.g.,
\begin{equation}
    K(\paramVec) = \frac{1}{\tau} \sum_{t=1}^{\tau} K(\paramVec; r_{m_t}) \,.
\end{equation}
More details on the matching procedure can be found in Ref.~\cite{thompson2009}.
This matching procedure requires derivative information, which can be obtained via finite differences.
However, we find that the least-squares approach is more accurate when using the Numerov method because the approach is derivative-free and thus does not give direct access to the derivative of the solution.

\subsubsection{Matrix Numerov method and its efficient implementation}
\label{sec:matrix_numerov}

We formulate the Numerov recurrence relation~\eqref{eq:numerov_recurrence}, combined with the initial values for $y_0$ and $y_1$ discussed in Sec.~\ref{sec:numerov_recurrence}, as the $(N-1) \times (N-1)$ linear system
\begin{equation} \label{eq:linear_system}
    \Amat(\paramVec) \vb{y}(\paramVec) = \vb{s}(\paramVec) \,
\end{equation}
and solve it for $\vb{y} = \{ y_2, y_3,\ldots, y_N\}$, with  
\begin{equation}\label{eq:matrix_numerov}
\begin{aligned}
\Amat &= 
\begin{bmatrix}
    \KnXi{2}{(1)} \\ 
    -2 \KnXi{2}{(-5)} & \KnXi{3}{(1)} \\ 
    \KnXi{2}{(1)}     & -2 \KnXi{3}{(-5)} & \KnXi{4}{(1)} \\ 
                  & \ddots        & \ddots    & \ddots \\
                  &               & \KnXi{N-2}{(1)} & -2 \KnXi{N-1}{(-5)} & \KnXi{N}{(1)} \\
    \end{bmatrix} \,, \\ 
\vb{s} &= 
\begin{bmatrix}
\left(2\KnXi{1}{(-5)} + \frac{1}{6}\delta_{\ell,1}\right) y_1 + \frac{h^2}{12} (s_2 + 10 s_1 + s_0)\\
- \KnXi{1}{(1)} y_1 + \frac{h^2}{12} (s_3 + 10 s_2 + s_1)\\
\vdots \\
\frac{h^2}{12} (s_N + 10 s_{N-1} + s_{N-2})
\end{bmatrix} \,.
\end{aligned} 
\end{equation}
Zero matrix elements are omitted for brevity throughout this paper.
We call this reformulation of the Numerov recurrence relation~\eqref{eq:numerov_recurrence} as a linear system the ``matrix Numerov method'' consistent with the physics literature~\cite{Pillai:2012,Jasper:2017,Miles:2018} on bound-state calculations.
Note that in Eq.~\eqref{eq:matrix_numerov}, we omitted the parameter dependence and that only the first two components of $\svec$ are populated when solving the inhomogeneous RSE.
Although we focus here on local potentials, we emphasize that the presented matrix Numerov method can also be applied to nonlocal potentials in coordinate space as discussed, e.g., in Refs.~\cite{Blanchon:2020ioa,Michel:2008sx}.

In the following, we discuss a computationally efficient implementation of the matrix Numerov method with a small memory and CPU footprint. 
We will take advantage of the sparsity of $\Amat$ and the affine decomposition of $V(r;\paramVec)$, which carries over to $g(r; \paramVec)$ and $s(r; \paramVec)$.
By construction, $\Amat$ is a lower triangular, banded matrix with bandwidth $B=3$.
It can brought into the matrix diagonal ordered form:\footnote{The matrix elements of $\Amat$ along the band and those of the matrix diagonal ordered form $\bar{\Amat}$ are related to one another by $\bar{\Amat}_{u + i - j +1, j} = \Amat_{i,j}$, where $u$ is the number of upper diagonals (i.e., superdiagonals).}
\begin{equation} \label{eq:Abar}
 \bar{\Amat} = 
            \begin{bmatrix} 
\KnXi{2}{(1)} & \KnXi{3}{(1)} & \cdots & \KnXi{N-1}{(1)} & \KnXi{N}{(1)}\\
-2\KnXi{2}{(-5)} & -2\KnXi{3}{(-5)} & \cdots & -2\KnXi{N-1}{(-5)} & *\\
\KnXi{2}{(1)} & \KnXi{3}{(1)} & \cdots & * & *\\
\end{bmatrix} \,,
\end{equation}
significantly reducing the overhead of storing zero matrix elements.
The matrix elements marked with $*$ are arbitrary and can be set to zero. 
They occur because there are more matrix elements along the diagonal than along the sub-diagonals and super-diagonals.
$\bar{\Amat}$ and $\vb{s}$ are then the input of
\texttt{SciPy}'s wrapper around LAPACK's efficient solver xGBSV for banded matrices, \texttt{linalg.solve\_banded()}~\cite{2020SciPy-NMeth}.

Next, we take advantage of the affine decompositions of $g(r;\paramVec)$ and $s(r;\paramVec)$, i.e., for the gridpoint $r_n$,\footnote{Note that the scalar $\paramVec_a$ corresponds to the $a$th component of $\paramVec$.}
\begin{subequations}    
\begin{align}
    g_n(\paramVec) &= \sum_{a=0}^{\ntheta} g_{na} \paramVec_a \,,\\
    g_{na} &= -2\mu \,\mathcal{V}_a(r_n) + \delta_{0,a} \left[ p^2 - \frac{l (l + 1)}{r_n^2} \right] \,,
    \intertext{and}
    s_n(\paramVec) &= \sum_{a=0}^{\ntheta} s_{na} \paramVec_a \,, \\
    s_{na} &= \zeta \, 2\mu \,\mathcal{V}_a(r_n) F(p r_n)\,,
\end{align}
\end{subequations}
which allows us to define values in an element-wise fashion ($i,j \in [1,2,\ldots,N-1]$)
\begin{subequations} \label{eq:matrix_numerov_A_affine}
\begin{align}
    \Amat_{ij} &= \sum_{a=0}^{n_\theta}  \vb*{\mathcal{A}}_{ija}\paramVec_a \,, \\
    \vb*{\mathcal{A}}_{ija} &= \delta_{a,0} \mathcal{D}_{i,j}(-2) + \frac{h^2}{12}  g_{(j+1)a} \, \mathcal{D}_{i,j}(10) \,, \\ 
    \mathcal{D}_{i,j}(x) &= \delta_{i,j} + x \delta_{j,i-1} + \delta_{j,i-2}  \,, 
\end{align}
\end{subequations}
and ($d \in [1,2,3]$ and $j \in [1,2,\ldots,N-1]$)
\begin{subequations} 
\begin{align}
    \bar{\Amat}_{dj}(\paramVec) &= \sum_{a=0}^{n_\theta} \bar{\vb*{\mathcal{A}}}_{dja} \paramVec_a  \,, \\
    \begin{split}
\bar{\vb*{\mathcal{A}}}_{dja} &= \delta_{a,0} \left(1 - 3\delta_{d,2} \right) \bar{\mathcal{D}}_{d,j}  \\
& \quad + \frac{h^2}{12} g_{(j+1)a}
\left(1 + 9 \delta_{d,2}\right) \bar{\mathcal{D}}_{d,j}   \,, 
\end{split}\\
\bar{\mathcal{D}}_{d,j} &= 1 -\delta _{3,d} \delta _{j,N-2}-(\delta _{2,d}+\delta _{3,d}) \delta_{j,N-1} \,.
\end{align}
\end{subequations}
Note that $\bar{\mathcal{D}}_{d,j}$ sets the irrelevant matrix elements in Eq.~\eqref{eq:Abar} to zero.
$\vb*{\mathcal{A}}$ and  $\bar{\vb*{\mathcal{A}}}$ are rank-3 tensors whose last (i.e., third) index corresponds to the model parameters.
Both can be prestored once upfront and then used to reconstruct the matrices $\Amat(\paramVec)$ and $\bar{\Amat}(\paramVec)$, respectively, for a given $\paramVec$ using efficient tensor multiplication.
This offline-online decomposition (even at the level of the FOM) makes the matrix Numerov method computationally very efficient in solving the RSE for different $\paramVec$.
Likewise, we express the right-hand side vector ($j \in [1,2,\ldots,N-1]$)
\begin{subequations} \label{eq:matrix_numerov_s_affine}
\begin{align}
    \vb{s}_j(\paramVec) &=  \sum_{a=0}^{n_\theta}  \vb{S}_{ja}\paramVec_a  \,,\\
    \begin{split}
    \vb{S}_{ja} &= 
    \delta_{a,0} \left[\left(2+\frac{1}{6}\delta_{\ell,1}  \right)  \delta_{j,1}  - \delta_{j,2} \right] y_1 \\ 
    & \quad + 
        \frac{h^2}{12} \biggl( -g_{1a} \left( 10\delta_{j,1} + \delta_{j,2}\right) y_1\\ 
    & \qquad\qquad + 
         s_{(j+1)a} + 10 s_{ja} + s_{(j-1)a} \biggr) \,,
     \end{split}
\end{align}
\end{subequations}
and prestore $\vb{S}_{ja}$ once upfront, allowing one to efficiently reconstruct $\vb{s}(\paramVec)$ for any given $\paramVec$ when solving the FOM.
The above tensors, which are defined element-wise, can be efficiently constructed using matrix-matrix and matrix-vector products. 
The output of this implementation still needs to be matched to the boundary condition~\eqref{eq:asymLimit} as described in Sec.~\ref{sec:numerov_recurrence}.
For completeness, we introduce in Appendix~\ref{app:other_emulators} the so-called ``all-at-once Numerov method,'' which directly imposes the asymptotic boundary condition for the scattering $T$-matrix element and for $\tau = 2$.
However, ROMs constructed from this variation of the matrix Numerov method are not guaranteed to preserve unitarity of the scattering $S$ matrix.

\subsubsection{Benchmarking the matrix Numerov method}
\label{sec:benchmark_numerov}

\begin{figure}[tb]
    \centering
\includegraphics[width=\columnwidth]{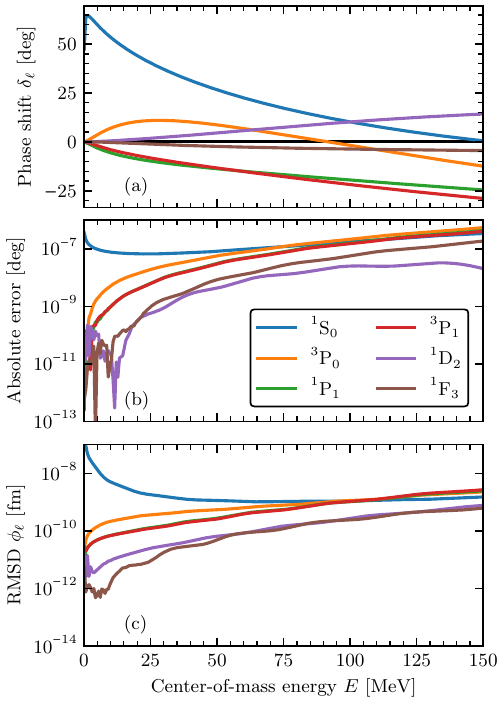}
    \caption{Benchmarks of the matrix Numerov method against the adaptive RK45 method, both applied to the inhomogeneous RSE~\eqref{eq:inhom_rse} subject to the boundary condition~\eqref{eq:asymLimit} through Eq.~\eqref{eq:def_chi}. 
    Panel~(a) shows the neutron-proton ($np$) scattering phase shifts in various uncoupled partial-wave channels (see the legend) for the GT+ chiral potential at \nTwoLO with cutoff $R_0 =1.0 \fm$ and spectral function cutoff $\Tilde{\Lambda} =1000 \MeV$.
    The corresponding absolute errors of the matrix Numerov method with respect to RK45 are depicted in panel~(b).
    Panel~(c) shows the corresponding root mean squared deviation (RMSD) of the scattered wave function $\chi(r)$.
    }
    \label{fig:numerov-benchmark}
\end{figure}

Next, we benchmark the matrix Numerov method against an adaptive ODE solver.
Figure~\ref{fig:numerov-benchmark}~(a) shows neutron-proton ($np$) phase shifts in several uncoupled partial-wave channels (see the legend) as a function of the center-of-mass energy $E$. 
The phase shifts are obtained by solving the inhomogeneous RSE~\eqref{eq:inhom_rse} with the matrix Numerov method applied to the local GT+ chiral potential at next-to-next-to leading order (\nTwoLO) with coordinate cutoff $R_0 = 1.0 \fm$ and spectral function cutoff $\Tilde{\Lambda} = 1000 \MeV$~\cite{Gezerlis:2014zia}.%
\footnote{We obtain similar accuracies for both the phase shifts and the full wave functions when solving the homogeneous RSE~\eqref{eq:rse}.}
The corresponding absolute errors with respect to the adaptive RK45 solver (for initial value problems) implemented in \texttt{SciPy}'s \texttt{integrate.solve\_ivp()}~\cite{2020SciPy-NMeth} are shown in Fig.~\ref{fig:numerov-benchmark}~(b). 
The accuracy of the phase shifts is correlated to the accuracy of the corresponding full wave functions, whose root mean squared deviations (RMSD) are depicted in Fig.~\ref{fig:numerov-benchmark}~(c).
We use here the standard spectroscopic notation for partial-wave channels ${}^{2s+1}\ell_j$, with the coupled spin $s = 0, 1$ and total angular momentum $j$, and naming convention for the angular momentum: $S \, (\ell = 0)$, $P \, (\ell = 1)$, $D \, (\ell = 2)$, and $F \, (\ell = 3)$.

To apply RK45, we express the second-order ODE~\eqref{eqn:formal-numerov} as the system of coupled first-order ODEs,
\begin{equation}
\begin{bmatrix} \vb{y}'_1(r;\paramVec)\\ \vb{y}'_2(r;\paramVec) \end{bmatrix} = 
\begin{bmatrix} \vb{y}_2(r;\paramVec) \\ f(r,\vb{y}_1(r;\paramVec);\paramVec) \end{bmatrix} \,,
\end{equation}
with the solution vector $\vb{y}(r;\paramVec) = \{y(r;\paramVec),y'(r;\paramVec)\}$.
The solver's relative and absolute tolerances are set to $\varepsilon_\text{rel} = \varepsilon_\text{abs}= 10^{-12}$ to obtain accurate results for the benchmark.
As initial values for RK45, we set $y'_0 = 0$  ($y'_0 = 1$) for solving the inhomogeneous (homogeneous) RSE and $y_0 = r_0 y'_0 / (l+1)$ following the discussion of the near-origin limit in Sec.~\ref{sec:numerov_recurrence}.
To avoid division by zero due to the centrifugal term for $\ell > 0$, we slightly shift grid points in the interval $r = [0, 12] \fm$ to $r_n = nh +\eta$, with $0 < \eta = 10^{-12} \ll h$ and $n = 0, 1, \ldots, N$.
Specifically, we use $N=10^3$ grid points with $\eta = 10^{-12}$ and the last $\tau = 25$ grid points to extract the phase shifts via the least-squares approach~\eqref{eq:lsq_matching}.
We find similar results when solving the homogeneous RSE~\eqref{eq:rse}.

In conclusion, Fig.~\ref{fig:numerov-benchmark} shows that phase shifts can be accurately extracted using the matrix Numerov method, with an accuracy of $\approx 10^{-7}$ degrees or better for $N=10^3$ grid points. 
The largest absolute error is in the \oneSzero channel at low energies, where the phase shift is the largest across the energies shown.

\section{Reduced-Order Model: Petrov-Galerkin Projections}
\label{sec:rom}

In this section, we examine two types of projection-based ROMs applied to the matrix Numerov method: 
the G-ROM results from a straightforward Galerkin projection of the linear system onto the subspace of (high-fidelity) solutions (see Sec.~\ref{sec:G-ROM});
the LSPG-ROM from a Petrov-Galerkin projection that minimizes the residual later defined in Eq.~\eqref{eq:residual} (see also Sec.~\ref{sec:lspg}). 
We implement efficient offline-online decompositions, exploiting affine decompositions of the emulator equations to improve the computational efficiency. 

The two emulators approximate the high-fidelity solution of the RSE at a given parameter vector $\paramVec$ as follows:
\begin{equation} \label{eq:rom_reduction}
    \yvec(\paramVec) \approx \ytilde(\paramVec) = 
    \Xmat \cvec(\paramVec) \,,
\end{equation}
where the $n_b$ column vectors of $\Xmat$ span the emulator's reduced space.
These column vectors can, and should, be made orthonormal for numerical stability such that $\Xmat^\dagger\Xmat = \vb{I}_{n_b}$.
Section~\ref{sec:snapshot_selection} provides a comprehensive discussion of how $\Xmat$ can be obtained. 
For now, we can assume $\Xmat$ is a given matrix of full rank.
The G-ROM and LSPG-ROM implement two different strategies to determine the coefficient vector $\cvec(\paramVec)$.

To keep the discussion applicable to linear systems in general,%
\footnote{In the general case, we only assume that $\Amat$ is invertible, so all eigenvalues are nonzero.}
we use the variable $\yvec(\paramVec)$ although we aim to solve the inhomogeneous RSE~\eqref{eq:inhom_rse} for the (unmatched) scattered wave function $\yvec(\paramVec) = \chi(\paramVec)$ from here on.
Solving for $\chi(\paramVec)$ has the numerical advantages discussed in Sec.~\ref{subsec:rse}. 
However, the main quantity of interest here is not the wavefunction; it is $K(\paramVec)$ and the corresponding $\delta(\paramVec)$, which are nonlinear functions of $\yvec(\paramVec)$.
Hence, we do not need to explicitly match the sequence obtained from the matrix Numerov method to the asymptotic limit parametrization~\eqref{eq:asymLimit}.
On the other hand, $a(\paramVec)$ and $b(\paramVec)$ are the least-squares solution of Eq.~\eqref{eq:lsq_matching} and as such depend only linearly on $\yvec(\paramVec)$:
\begin{equation} \label{eq:lsq_sol_a_b}
    \begin{bmatrix} a(\paramVec) \\ b(\paramVec) \end{bmatrix} =
\Dmat^+ \vb{F}^\dagger \yvec(\paramVec) +     \begin{bmatrix} \zeta p \\ 0 \end{bmatrix} \,,
\end{equation}
where the superscript $+$ denotes the pseudo-inverse,
\begin{equation}
\Dmat = 
    \frac{1}{p} 
    \begin{bmatrix}
        F(p r_{m_{1}})  & G(p r_{m_{1}}) \\
        F(p r_{m_{2}})  & G(p r_{m_{2}}) \\
        \vdots & \vdots \\
        F(p r_{m_{\tau}})  & G(p r_{m_{\tau}}) 
    \end{bmatrix}
\end{equation}
is the design matrix in Eq.~\eqref{eq:lsq_matching}, and $\vb{F}$ is a $(N-1) \times \tau$ matrix with components $\vb{F}_{it} = \delta_{i,m_t}$. 
As in Sec.~\ref{sec:fom}, $\zeta = 0$ ($\zeta = 1$) when solving the homogeneous (inhomogeneous) RSE.
Equation~\eqref{eq:lsq_sol_a_b} can be straightforwardly obtained by writing Eq.~\eqref{eq:lsq_matching} as
\begin{equation}
\Dmat \begin{bmatrix} a(\paramVec) - \zeta p \\ b(\paramVec) \end{bmatrix} = \vb{F}^\dagger \yvec(\paramVec) \,.
\end{equation}
Using Eq.~\eqref{eq:rom_reduction}, $a(\paramVec)$ and $b(\paramVec)$ are emulated by
\begin{equation} \label{eq:lsq_sol_a_b_tilde}
    \begin{bmatrix} \tilde{a}(\paramVec) \\ \tilde{b}(\paramVec) \end{bmatrix} =
\left( \Dmat^+  \vb{F}^\dagger \Xmat \right) \cvec(\paramVec) + \begin{bmatrix} \zeta p\\ 0 \end{bmatrix} \,,
\end{equation}
where the $2 \times n_b$ matrix $\Dmat^+  \vb{F}^\dagger \Xmat $ can be precomputed in the offline stage.
Each matrix column is determined by a linear combination of the $a(\paramVec)$ and $b(\paramVec)$ of the FOM calculations performed to train the emulator.
Finally, the corresponding approximations for $K(\paramVec)$ and $\delta(\paramVec)$ are given by Eqs.~\eqref{eq:K_delta_a_b} with $a(\paramVec)$ and $b(\paramVec)$ replaced by $\tilde{a}(\paramVec)$ and $\tilde{b}(\paramVec)$, respectively.

\subsection{Galerkin ROM}
\label{sec:G-ROM}

The G-ROM determines the coefficient vector $\cvec(\paramVec)$ in Eq.~\eqref{eq:rom_reduction} by Galerkin projection, i.e., by requiring that the residual
\begin{equation}\label{eq:residual}
    \residual(\paramVec) = \svec(\paramVec) - \Amat(\paramVec) \ytilde(\paramVec)
\end{equation}
is orthogonal to the columns of $\Xmat$.
Thus, $\Xmat^\dagger \residual(\paramVec) = 0$, together with Eq.~\eqref{eq:rom_reduction}, become the G-ROM equations:
 \begin{multline} \label{eq:G-ROM_lin_system}
        \tAmat(\paramVec) \cvec(\paramVec) = \tsvec(\paramVec) \\ 
        \text{with} \quad 
    \tAmat(\paramVec) = \Xmat^\dagger \Amat(\paramVec) \Xmat \,, \quad 
    \tsvec(\paramVec) = \Xmat^\dagger \svec(\paramVec) \,.
 \end{multline}
Notice that $\Amat$ is an $(N-1) \times (N-1)$ matrix, while $\tAmat$ is an $n_b \times n_b$ matrix, with $n_b \ll N$.
Correspondingly, $\svec$ is a length--$(N-1)$ vector while $\tsvec$ is a length-$n_b$ vector.
To approximate the inhomogeneous RSE~\eqref{eq:inhom_rse} for an arbitrary parameter vector $\paramVec$ using the matrix Numerov method, we solve the reduced system~\eqref{eq:G-ROM_lin_system} for $\cvec(\paramVec)$ instead of the full-order system~\eqref{eq:linear_system} [with Eqs.~\eqref{eq:matrix_numerov_A_affine} and~\eqref{eq:matrix_numerov_s_affine}] for $\yvec(\paramVec)$.

Next, we take advantage of the affine decompositions of $\Amat$ (see Eq.~\eqref{eq:matrix_numerov_A_affine}) and $\svec$ (see Eq.~\eqref{eq:matrix_numerov_s_affine}), which carry over to the G-ROM~\eqref{eq:G-ROM_lin_system}:
\begin{align}
\tAmat_{uv}(\paramVec) &= \sum_{a=0}^{n_\theta} \left[\sum \limits_{i,j=1}^{N-1} \Xmat_{iu}^* \vb*{\mathcal{A}}_{ija} \Xmat_{jv}\right]_{uva} \paramVec_a \,, \label{eq:Atilde_ij}\\
\begin{split}
\tsvec_u(\paramVec) &= \sum_{a=0}^{n_\theta}  \left[ \sum_{i=1}^{N-1} \Xmat^*_{iu} \vb{S}_{ia} \right]_{ua} \paramVec_a \,.
\end{split}\label{eq:stilde_ij}
\end{align}
Each of the tensors in brackets $[\ldots]$ can be prestored in the emulator's offline stage, from which $\tAmat$ and $\tsvec$ can be efficiently reconstructed in the emulator's online stage. 
The size reduction of the linear system to be solved (i.e., $n_b \ll N$) together with an offline-online decomposition allows for constructing fast \& accurate emulators.

In summary, the G-ROM workflow is as follows:
\begin{itemize}
  \item In the offline stage, we prestore the tensors in the brackets $[\ldots]$ of Eqs.~\eqref{eq:Atilde_ij} and~\eqref{eq:stilde_ij}.
  \item In the online stage, we reconstruct $\tAmat(\paramVec)$ and $\tsvec(\paramVec) $ using Eqs.~\eqref{eq:Atilde_ij} and~\eqref{eq:stilde_ij}, respectively, from these prestored tensors and solve the ROM equation~\eqref{eq:G-ROM_lin_system} for the coefficient vector $\cvec(\paramVec)$. 
  \item Equation~\eqref{eq:rom_reduction} then provides the ROM approximation to the high-fidelity solution $\yvec(\paramVec)$.
\end{itemize}
%

\subsection{Least-squares Petrov-Galerkin ROM}
\label{sec:lspg}

The idea behind least-squares Petrov-Galerkin (LSPG) projection is to compute $\cvec(\paramVec)$ such that the norm of $\residual(\paramVec)$ in Eq.~\eqref{eq:residual} is minimized~\cite{Quarteroni:218966}.
This leads to solving the least-squares problem
\begin{equation}\label{eq:lspg_ls_prob}
    \Amat(\paramVec) \Xmat \cvec(\paramVec) = \svec(\paramVec) \,.
\end{equation}
Considering the normal equations
\begin{equation}
    \Xmat^\dagger \Amat^\dagger(\paramVec) \Amat(\paramVec) \Xmat \cvec(\paramVec)
    = \Xmat^\dagger \Amat^\dagger(\paramVec) \svec(\paramVec) \,,
\end{equation}
we see that it can be interpreted as a Petrov-Galerkin projection of Eq.~\eqref{eq:linear_system} with right projection by $\Xmat$ and left projection by $\Amat(\paramVec) \Xmat$ (which motivates the name LSPG).
Alternatively, it can be interpreted as a Galerkin projection of Eq.~\eqref{eq:linear_system} after left-multiplication by $\Amat^\dagger(\paramVec)$:
\begin{equation}
    \Amat^\dagger(\paramVec) \Amat(\paramVec) \yvec(\paramVec)
    = \Amat^\dagger(\paramVec) \svec(\paramVec) \,,
\end{equation}
combined with the right and left projections by $\Xmat$.

The benefit of the LSPG-ROM compared to the G-ROM is that the matrix $\Amat^\dagger \Amat$ is Hermitian positive-definite (h.p.d.).
Thus, since $\vb{X}$ is of full rank, $\Xmat^\dagger \Amat^\dagger \Amat \Xmat$ is also h.p.d., and, in particular, it is invertible (albeit it may still be poorly conditioned).
Hence, it is expected to be significantly less susceptible to spurious singularities that occur at points in the parameter space where no (unique) solutions to the emulator equations exist due to singular emulator matrices.
We identify these spurious singularities as Kohn anomalies, which are well-known in the context of the Kohn variational principle (KVP).
More information on Kohn anomalies and their mitigation can be found in Ref.~\cite{Drischler:2021qoy}.

It is possible to approach the offline-online decomposition as with the G-ROM in Sec.~\ref{sec:G-ROM}.
The issue is that the affine decompositions of $\Amat^\dagger \Amat$ and $\Amat^\dagger \svec$ have about $n_{\theta}^2$ terms, which would make the online phase more computationally demanding.
Furthermore, it may be numerically beneficial to avoid solving the least-squares problem via normal equations because the condition number of $\Amat^\dagger \Amat$ is equal to the squared condition number of $\Amat$.
Instead, based on Ref.~\cite{Buhr:2014}, we project Eq.~\eqref{eq:lspg_ls_prob} onto the subspace of residuals as explained in the following.

Let us note that the residual~\eqref{eq:residual} can be written as:
\begin{align}  
\residual(\paramVec)
&= \sum_{a=0} \left[  \svec^{(a)} - \Amat^{(a)} \Xmat \cvec(\paramVec) \right] \paramVec_a \,,
\label{eq:residual_affine}\end{align}  
where we used the short-hand notation for the column vector $\svec^{(a)}$ with components $(\svec^{(a)})_{i} = \vb{S}_{ia}$ and the matrix $\vb{A}^{(a)}$ with components $(\vb{A}^{(a)})_{ij}= \{\vb{\mathcal{A}}_{ija}\}_{ij}$.
Note that $\residual(\paramVec) = \vb{0} $ if $\ytilde(\paramVec) = \vb{y}(\paramVec)$.
From Eq.~\eqref{eq:residual_affine}, we can see that the residual $\residual(\paramVec)$ for \emph{any} parameter value $\paramVec$ lives in the subspace spanned by the column vectors of $\vb{B}^{(a)} = \Amat^{(a)} \Xmat$ ($N \times n_b$ matrices) and the right-hand side vectors $\vb{s}^{(a)}$ (length-$N$) for all $a \in [0,1,\ldots,n_\theta]$.
We stack these vectors and matrices horizontally to obtain the $N \times (n_\theta+1)(n_b+1)$ matrix:
\begin{equation} \label{eq:def_Ymat}
    \Ymat = \begin{bmatrix} \vb{B}^{(0)} & \vb{B}^{(1)} & \cdots & \vb{B}^{(n_\theta)} & \vb{s}^{(0)} & \vb{s}^{(1)} & \cdots & \vb{s}^{(n_\theta)}\end{bmatrix} \,.
\end{equation}
We then orthonormalize and truncate $\Ymat$'s column vectors, which provide the projection basis, following the procedure detailed in Sec.~\ref{sec:pod}. 
It then has dimensions $N\times n_{\Ymat}$, with $n_{\Ymat} \leqslant (n_\theta+1)(n_b+1)$.
The matrix $\Ymat \Ymat^\dagger$ is an orthogonal projector onto the (approximate) subspace of residuals of our ROM, which will also be an important property when deriving emulator error estimates in Sec.~\ref{sec:greedy_algorithm}.

Thus, our LSPG-ROM is the projected least-squares problem
\begin{multline} \label{eq:lspg_lin_system}
    \tAmat(\paramVec) \cvec(\paramVec) = \tsvec(\paramVec) \\
    \text{with} \quad 
    \tAmat(\paramVec) = \Ymat^\dagger \Amat(\paramVec) \Xmat \,, \quad 
    \tsvec(\paramVec) = \Ymat^\dagger \svec(\paramVec) \,,
\end{multline}
and $\cvec(\paramVec)$ is computed using the least-squares solver implemented in \texttt{SciPy}'s \texttt{linalg.lstsq()}. 
It can be checked that Eqs.~\eqref{eq:lspg_ls_prob} and~\eqref{eq:lspg_lin_system} have the same solution if the truncation in $\Ymat$ was exact (otherwise, the solutions are approximately equal).

Now, similar to the G-ROM, we can take advantage of the affine decompositions (with $w \in [1,2,\ldots,n_{\Ymat}]$):
\begin{align}
\tAmat_{wu}(\paramVec) &= \sum_{a=0}^{n_\theta} \left[\sum \limits_{i,j=1}^{N-1} \Ymat_{iw}^* \vb*{\mathcal{A}}_{ija} \Xmat_{ju}\right]_{wua} \paramVec_a \,, \label{eq:Atilde_ij_lspg}\\
\begin{split}
\tsvec_w(\paramVec) &= \sum_{a=0}^{n_\theta}  \left[ \sum_{i=1}^{N-1} \Ymat^*_{iw} \vb{S}_{ia} \right]_{wa} \paramVec_a \,.
\end{split}\label{eq:stilde_ij_lspg}
\end{align}
Again, each of the tensors in brackets $[\ldots]$ can be prestored once in the emulator's offline stage, from which $\tAmat$ and $\tsvec$ can be efficiently reconstructed in the emulator's online stage. 
Hence, the LSPG-ROM workflow in the online stage is similar to the one for the G-ROM, with the difference that Eqs.~\eqref{eq:Atilde_ij_lspg} and~\eqref{eq:stilde_ij_lspg} determine the reduced linear system to be solved for $\cvec$.

In summary, the LSPG-ROM workflow is as follows:
\begin{itemize}
  \item In the offline stage, we prestore the tensors in the brackets $[\ldots]$ of Eqs.~\eqref{eq:Atilde_ij_lspg} and~\eqref{eq:stilde_ij_lspg}.
  \item In the online stage, we reconstruct $\tAmat(\paramVec)$ and $\tsvec(\paramVec) $ using Eqs.~\eqref{eq:Atilde_ij_lspg} and~\eqref{eq:stilde_ij_lspg}, respectively, from these prestored tensors and solve the ROM equation~\eqref{eq:lspg_lin_system} for the coefficient vector $\cvec(\paramVec)$. 
  \item Equation~\eqref{eq:rom_reduction} then provides the ROM approximation to the high-fidelity solution $\yvec(\paramVec)$.
\end{itemize}

\section{Snapshot selection}
\label{sec:snapshot_selection}

In this section, we discuss two common approaches to selecting the snapshot locations used to construct the ROM.  
These methods are the POD 
and greedy algorithm (see Fig.~\ref{fig:pod-vs-greedy} for an illustration).  
We also discuss estimating emulator errors in the wave functions and propagating them to scattering phase shifts.

\subsection{Proper orthogonal decomposition}
\label{sec:pod}

\begin{figure}[tb]
    \centering
    \includegraphics[width=\columnwidth]{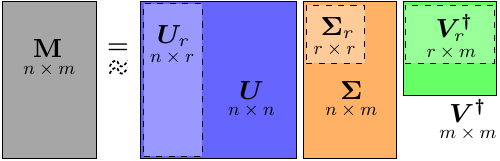}
    \caption{
    Illustrations of the SVD (solid outline) and truncated SVD (dashed outline) applied to a real or complex matrix $\Mmat$: $\Umat$ and $\Vmat$ are unitary matrices (e.g., $\Umat^\dagger \Umat = \Umat\Umat^\dagger = \vb{1}$) containing the singular vectors; $\vb*{\Sigma}$ is a diagonal matrix containing the singular values in descending order. 
    The truncated SVD truncates the singular vectors corresponding to the $r$ smallest singular values, resulting in a best possible rank-$r$ approximation (in the Frobenius norm) to the original $\Mmat$, also called low-rank approximation.
    The POD is based on a truncated SVD of the snapshot matrix, taking only the (orthonormal) left singular vectors associated with the $r$ largest singular values. 
    The matrix dimensions are annotated.
    }
    \label{fig:illustration_svd}
\end{figure}

The POD approach to snapshot selection first samples the parameter space, e.g., using a space-filling algorithm. 
These random samples are illustrated as the red dots in the left panel of Fig.~\ref{fig:pod-vs-greedy}.
High-fidelity calculations are then carried out for each sample and stacked as the columns of a matrix $\Mmat$. 
Next, one orthonormalizes and compresses these column vectors by applying a truncated singular value decomposition (SVD) to $\Mmat = \Umat\vb*{\Sigma}\Vmat^\dagger$. 

As illustrated in Fig.~\ref{fig:illustration_svd}, the SVD is a matrix decomposition of real or complex matrices in terms of two unitary matrices, $\Umat$ and $\Vmat$, and a diagonal matrix $\vb*{\Sigma}$ with nonnegative diagonal entries in decreasing order.
The orthonormal column vectors of $\Umat$ ($\Vmat$) are called the left (right) singular vectors, and the diagonal entries of $\vb*{\Sigma}$ are called the singular values and denoted with $\sigma_i(\Mmat)$.
The span of the left singular vectors corresponding to the positive singular values equals the span of $\Mmat$'s column vectors.
Furthermore, truncating the SVD to $\Umat_r \vb*{\Sigma}_r \Vmat_r^\dagger$ gives a best rank-$r$ approximation of $\Mmat$
for any $r \leqslant \min(m, n)$,%
\footnote{Here, the generic number of rows $n$ corresponds to $N$ while the generic number of columns $m$ corresponds to the number of initial high-fidelity calculations, so in our case, $m \ll n$.
Furthermore, $r$ corresponds to the number of basis vectors of the emulator, $n_b$.}
where $\Umat_r$ and $\Vmat_r$ consist of the first $r$ columns of $\Umat$ and $\Vmat$, respectively, and $\vb*{\Sigma}_r$ is the upper-left $r \times r$ submatrix of $\vb*{\Sigma}$.%
\footnote{We follow here the standard notation in mathematics that uses $r$ for the rank of a matrix, which is not to be confused with the radius in this section.}
In particular, by the Eckart-Young-Mirsky theorem, we have that $\norm{\Mmat - \Umat_r \vb*{\Sigma}_r \Vmat_r^\dagger}_2 = \sigma_{r+1}(\Mmat)$ and $\norm{\Mmat - \Umat_r \vb*{\Sigma}_r \Vmat_r^\dagger}_F^2 = \sum_{i = r + 1}^{\min(m, n)} \sigma_i^2(\Mmat)$, where $\norm{\bullet}_2$ is the matrix $2$-norm (also equal to the largest singular value) and $\norm{\bullet}_F$ is the Frobenius norm.

The POD then sets the emulator's basis matrix $\Xmat$ as $\Umat_r$ for some $r$ (the left singular vectors of the snapshot matrix $\Mmat$ are also called POD modes).
One may include all left singular vectors whose singular values fulfill the condition $\sigma_i(\Mmat) \geqslant \eta \, \sigma_\mathrm{max}(\Mmat)$, where $\eta$ is a small truncation tolerance and $\sigma_\mathrm{max}(\Mmat)$ the maximum singular value of $\Mmat$.
Another standard option would be considering all left singular vectors whose singular values fulfill
\begin{equation}
    \frac{\sum_{i=1}^{r} \sigma_i^2(\Mmat)}{\sum_{i=1}^{\min(m,n)}\sigma_i^2(\Mmat)}
    \leqslant 1 - \eta \,.
\end{equation}
This ratio is sometimes referred to as the amount of preserved variance. 
Assuming $\Xmat$ contains the FOM calculations (i.e., before applying the POD) and $\vb{U}_r$ the corresponding dominant POD modes (i.e., after applying the POD), then the coefficients obtained with these two emulator bases are related to each other as $\cvec' \simeq ( \vb{V}_r \vb*{\Sigma}_{r}^{-1} ) \cvec$ and $\vb{U}_r \cvec \simeq \Xmat \cvec'$.
The equal sign applies if, and only if, the POD truncates no or only zero singular values.

The strength of the described POD approach is that it contains much information on high-fidelity solutions across the parameter space through space-filling sampling.
However, this high level of information comes at the computational expense of performing many high-fidelity calculations in the emulator's offline stage, which can be prohibitively slow.
Furthermore, the large number of initial high-fidelity calculations may contain superfluous information, indicated by rapidly decreasing singular values and thus high compression rates with $r \ll \min(m,n)$.

\subsection{Greedy algorithm}
\label{sec:greedy_algorithm}

Given an initial (small) emulator basis, the greedy algorithm improves the emulator's accuracy by iteratively adding snapshots in locations of the maximum estimated error, until the requested accuracy goal is achieved.
Estimating the emulator's errors across the parameter space is therefore required by this approach to snapshot selection (see Sec.~\ref{sec:error_estimates}). 
The initial emulator basis could be 
formed using the POD approach described in Sec.~\ref{sec:pod} or otherwise determined.
This algorithm is \emph{greedy} because it makes locally optimal choices in reducing the emulator error (see Sec.~\ref{sec:greedy_algorithm}). 
However, these choices may not lead to globally optimal emulator bases (see, e.g., Ref.~\cite{Ohlberger:2016}, for results on quasi-optimality).

\subsubsection{Error estimates and bounds}
\label{sec:error_estimates}

The high-fidelity solution vector $\yvec(\paramVec)$ and its ROM approximation in Eq.~\eqref{eq:rom_reduction} differ by the error\footnote{We use the standard nomenclature in the model reduction community: If $\Amat \yvec=\vb{s}$ and $\ytilde$ is an approximate solution, then $\rvec=\svec-\Amat \ytilde$ is called the residual and $\vb{e}=\yvec-\ytilde$ is called the error. The two are related by $\Amat \vb{e} = \rvec$, which is used in the error estimator.}
\begin{equation} \label{eq:rom_error}
    \exactErrorVec(\paramVec) = \yvec(\paramVec) - \ytilde(\paramVec)  \,,
\end{equation}
which vanishes if and only if $\yvec(\paramVec)$ is in the column space of $\Xmat$; e.g., at a snapshot location in the parameter space.
In the following, we aim to efficiently construct an estimator for the norm of the error~\eqref{eq:rom_error} without performing expensive high-fidelity calculations.

In lieu of the in-practice unknown error~\eqref{eq:rom_error}, we use the residual of the high-fidelity linear system for the ROM solution in Eq.~\eqref{eq:residual} as a computationally efficient proxy. 
Although the residual~\eqref{eq:residual} is generally not an approximation for the exact error~\eqref{eq:rom_error}, it can be used to derive bounds on $\norm{\exactErrorVec(\paramVec)}$ if one has access to the extremal singular values of $\Amat$, $\sigma_\mathrm{min}(\Amat)$ and $\sigma_\mathrm{max}(\Amat)$, respectively. 
Specifically, one finds rigorous theoretical bounds in the spectral norm (i.e., the $\ell^2$ or Euclidean norm for vectors)\footnote{The derivation uses a Cauchy–Schwarz-like inequality and the fact that $\norm{\Amat}_2 = \sigma_\mathrm{max}(\Amat)$ in the spectral norm. 
Likewise, $\norm{\Amat^{-1}}_2 = \sigma^{-1}_\mathrm{min}(\Amat)$.
The spectral (matrix) norm coincides with the Euclidean norm for vectors. We omit the subscript $2$ that indicates the spectral norm.}
\begin{equation} \label{eq:error_bounds}
    \frac{\norm{\romResidual(\paramVec)}}{\sigma_\mathrm{max}(\Amat(\paramVec))} 
    \leqslant \norm{\exactErrorVec(\paramVec)} 
    \leqslant \frac{\norm{\romResidual(\paramVec)}}{\sigma_\mathrm{min}(\Amat(\paramVec))} \,,
\end{equation}
where the upper bound is mainly of interest because it may serve as a conservative error estimate.
Note that we assume that $\Amat(\paramVec)$ is invertible so that $\sigma_\mathrm{min}^{-1}(\Amat(\paramVec))$ is defined for all parameter values $\paramVec$.
However, these extremal singular values might be expensive to compute. 

Several ways to construct an error estimator exist.
A conservative approach is to use the upper bound in Eq.~\eqref{eq:error_bounds} as the (largest possible) error estimate, assuming $\sigma_\mathrm{min}(\Amat(\paramVec))$ is known or can be efficiently estimated, e.g., using the successive constraint method (SCM). 
The SCM gives a lower bound for $\sigma_\mathrm{min}(\Amat(\paramVec))$ for a new parameter vector $\paramVec$ based on previously computed smallest singular values for other parameter vectors.
For our numerical examples, the SCM needed hundreds of samples to obtain a sufficiently tight lower bound (compared to less than ten samples needed for the reduced basis),
therefore, more work is needed to render the SCM efficient enough for our applications.
Another less rigorous approach is constructing two emulators with different basis sizes and using the difference between the emulator predictions to estimate the true error. 
This approach is called the hierarchical posteriori error estimation (HPEE).
In practice, we find that $\norm{\romResidual(\paramVec)}$ and $\norm{\exactErrorVec(\paramVec)}$ are approximately proportional to each other (see Sec.~\ref{sec:results}).
Under this assumption, which should be validated in each application and was previously investigated in Ref.~\cite{Sarkar:2021fpz}, the location of maximum error for the greedy algorithm can be approximated efficiently without estimating the smallest singular value.

To compute the residual in the full space~\eqref{eq:residual} efficiently, let us define the projected residual
(based on Ref.~\cite{Buhr:2014} and discussed in Sec.~\ref{sec:lspg})
\begin{equation}  \label{eq:rom_residual_Y}
    \romResidualY(\paramVec) \equiv \Ymat^\dagger \romResidual(\paramVec) = \Ymat^\dagger \svec(\paramVec) - \Ymat^\dagger \Amat(\paramVec) \ytilde (\paramVec) \,.
\end{equation}  
It is a vector of length $(n_\theta+1)(n_b+1)$, living in a semi-reduced space, not the reduced or full space if it has less than $N$ components. 
As discussed in Sec.~\ref{sec:lspg}, we construct the column vectors of $\Ymat$ to be orthonormal, so that $\vb{P} = \Ymat \Ymat^\dagger$ is an orthogonal projector onto the space of the residuals~\eqref{eq:residual}, with $\vb{P} \romResidual(\paramVec) = \romResidual(\paramVec)$ for all $\romResidual(\paramVec)$. 
This property implies that the residual vector in the full space~\eqref{eq:residual} can be computed exactly and computationally more efficiently in the semi-reduced space:
\begin{equation}
     \norm{\romResidual(\paramVec)} = \norm{\romResidualY(\paramVec)} \,.
\end{equation}
In the emulator's online stage, we can further reconstruct $\romResidualY(\paramVec)$ from tensors precomputed in the offline stage:
\begin{equation} \label{eq:romResidualY_affine}
\begin{split}
      (\romResidualY)_w(\paramVec) &= 
 \sum_{a=0}^{n_\theta}  \left[ \sum_{i=1}^{N-1} \Ymat^{*}_{iw}  \vb{S}_{ia} \right]_{wa} \paramVec_a \\
& \quad - \sum_{a=0}^{n_\theta} \sum_{u=0}^{n_b} \left[\sum_{i,j=1}^{N-1} \Ymat_{iw}^* \vb*{\mathcal{A}}_{ija} \Xmat_{ju} \right]_{wua} \paramVec_a \cvec_u(\paramVec) \,,
\end{split}
\end{equation}
where the terms in the brackets can be prestored.
Once $\romResidualY(\paramVec)$ is reconstructed in the online stage, we compute its Euclidean norm, guaranteeing that its magnitude is nonnegative, as expected.

\begin{figure*}[tb]
    \centering
    \includegraphics[width=\textwidth]{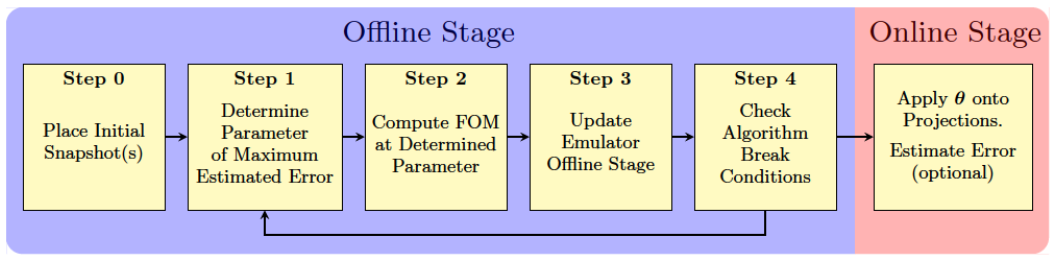}
    \caption{%
    Flowchart depicting the process of constructing an emulator basis using the greedy algorithm developed in this work. 
    The greedy algorithm iteratively improves an initial emulator basis in the offline stage. 
    Once the greedy algorithm is terminated, the emulator can be used in the online stage, similar to the POD (with a different emulator basis).
    See the main text for details.}
    \label{fig:greedy-flowchart}
\end{figure*}

Alternatively, we investigated reconstructing the magnitude-squared of the residual~\eqref{eq:residual} directly from the affine decomposition of
\begin{equation} \label{eq:residual_norm2}
\begin{split}
    \norm{\romResidual(\paramVec)}^2 
&= \ytilde^\dagger(\paramVec) \Amat^\dagger(\paramVec) \Amat(\paramVec) \ytilde(\paramVec)  \\
& \quad -  2 \Re \left[ \ytilde^\dagger(\paramVec) \Amat^\dagger(\paramVec) \svec(\paramVec) \right] \\
& \quad + \svec^\dagger(\paramVec) \svec(\paramVec) \,,
\end{split}
\end{equation}
where $\Re(\bullet)$ denotes the real part of the complex-valued argument.
Appendix~\ref{app:prestore_scalar_residual} discusses the details of this affine decomposition.
However, the reconstruction is susceptible to round-off errors in finite-precision arithmetic due to the cancellation in Eq.~\eqref{eq:residual_norm2}.
These numerical artifacts may violate the strict non-negativeness of the reconstructed $\norm{\romResidual(\paramVec)}^2 $.
Hence, we use Eqs.~\eqref{eq:rom_residual_Y} through~\eqref{eq:romResidualY_affine} to construct an efficient offline-online decomposition.

\subsubsection{Error propagation to phase shifts}
\label{sec:UQ_propagation}

Next, we discuss the propagation of the ROM error~\eqref{eq:rom_error} to phase shifts, which is the main quantity of interest here. 
To this end, we expand, at a given~$\paramVec$, the error in $\delta(\paramVec)$ up to first order in the asymptotic limit coefficients $a(\paramVec)$ and $b(\paramVec)$. 
Its norm can be bounded via the Cauchy-Schwarz inequality, leading to
\begin{align}
    \Delta(\paramVec) = \left\lvert \delta(\paramVec) - \tilde{\delta}(\paramVec) \right\rvert &\lesssim \frac{180}{\pi} \frac{\norm{ \begin{bmatrix} 
    a(\paramVec) - \tilde{a}(\paramVec) \\ b(\paramVec) - \tilde{b}(\paramVec) 
    \end{bmatrix}} }{\sqrt{|\tilde{a}(\paramVec)|^2 + |\tilde{b}(\paramVec)|^2}}  \,, \label{eq:norm_error_delta}
\end{align}
where the denominator is due to the norm of the Jacobian of Eq.~\eqref{eq:delta_from_K_a_b} and the phase shifts are measured in degrees rather than radians.
These norms are spectral norms, as before.
Using Eqs.~\eqref{eq:lsq_sol_a_b} and~\eqref{eq:lsq_sol_a_b_tilde}, we likewise derive:
\begin{equation} \label{eq:lsq_sol_a_b_error}
   \norm{ \begin{bmatrix} 
    a(\paramVec) - \tilde{a}(\paramVec) \\ b(\paramVec) - \tilde{b}(\paramVec) 
    \end{bmatrix} }
    \leqslant 
    \norm{\Dmat^+  \vb{F}^\dagger} \; \norm{\exactErrorVec(\paramVec)} \,.
\end{equation}
The first term on the right-hand of Eq.~\eqref{eq:lsq_sol_a_b_error} can be prestored in the online stage because it depends only on the scattering energy and the number of points used for solving the least-squares problem~\eqref{eq:lsq_matching}; 
the second term can be bounded using Eq.~\eqref{eq:error_bounds} or estimated as described in the next subsection.
Equations~\eqref{eq:norm_error_delta} and~\eqref{eq:lsq_sol_a_b_error} together provide an upper bound on $\Delta(\paramVec)$, which could be treated as a (conservative) symmetric error estimate of the phase shift, $\delta(\paramVec) \pm \Delta(\paramVec)$.
Future work may explore statistically more rigorous discrepancy models based on this error estimate for the phase shifts.

\subsubsection{Iterative improvement of the emulator basis}
\label{sec:iterative_improvement}

The workflow of the greedy algorithm is illustrated in Fig.~\ref{fig:greedy-flowchart}. 
It can be summarized as follows.

In Step~0, one has to decide on the number of initial snapshots and where they are placed in the parameter space. 
Each snapshot requires a high-fidelity calculation. 
This step needs to be completed only once.
For example, the greedy algorithm could start with a single snapshot at the center of the parameter space to train the emulator with the fewest high-fidelity solutions possible. 
However, with only one snapshot in the emulator's basis, the accuracy of the error estimate may be initially relatively low.
One could also place snapshots more physics-informedly, e.g., in locations where bound states or resonances are expected, or one could use LHS with a few sampling points.
To improve numerical stability, we recommend that the (initial) emulator basis be orthonormalized, e.g., via the $QR$ decomposition of the initial emulator basis.

In Step~1, one determines the location of the maximum estimated error in the parameter space.
This optimization problem is solved by emulating high-fidelity solutions.
Here, we densely sample the parameter space using LHS once in Step~0 to generate a pool of a few thousand candidate snapshots, rendering the continuous optimization problem discrete.
The high-fidelity solutions for each candidate snapshot are then emulated, and the pool member with the maximum error is identified.

In Step~2, one computes the high-fidelity solution at the location determined in the previous step.
The solution is orthonormalized with respect to the vectors already in the emulator basis.
We compute the updated $QR$ factorization [e.g., using \texttt{SciPy}'s \texttt{qr\_insert()} function] to efficiently orthornomalize the candidate snapshot with respect to the current emulator basis, which is already orthonormal.

In Step~3, one updates the emulator's offline stage.
All tensors from which the emulator equation is reconstructed in the online stage must be updated to reflect the new emulator basis. 

In Step~4, one checks the break condition and moves on to the next greedy iteration.
Repeat Steps~1 through~4 until a break condition, such as the requested emulator accuracy or the maximum number of iterations, is reached. 
If desired, the greedy algorithm can be resumed later to improve the emulator basis further.

Assuming that $\norm{\romResidual(\paramVec)}$ and $\norm{\exactErrorVec(\paramVec)}$ are approximatively proportional to each other across the parameter space, one can calibrate the error estimator discussed in Sec.~\ref{sec:error_estimates} such that $\kappa \norm{\romResidual(\paramVec)} \approx \norm{\exactErrorVec(\paramVec)}$.
To this end, one has to perform only one high-fidelity calculation at a point in the parameter space $\bar{\paramVec}$ where the emulator error is nonzero and then compute the scaling factor  $\kappa = \norm{\exactErrorVec(\bar{\paramVec})} / \norm{\romResidual(\bar{\paramVec})}$.
The calibration does not affect the location of the maximum estimated error.
In practice, after the break condition is met in Step~4, we suggest letting the greedy algorithm repeat Steps~1 and~2 one more time and using the high-fidelity solution obtained in Step~2 to calibrate the error estimator as described.
The output of the calibrated error estimator could then be added to the overall UQ of the calculation, which may include EFT truncation errors~\cite{Wesolowski:2021cni}.
Finally, we note that although the greedy algorithm described above makes decisions based on the errors of the emulated wave functions, reducing those errors also reduces the error estimate of the phase shift through Eqs.~\eqref{eq:norm_error_delta} and~\eqref{eq:lsq_sol_a_b_error}, as expected.

\section{Results and Discussion}
\label{sec:results}

In this section, we present our main results.
We discuss the simple Minnesota potential as a test case in Sec.~\ref{sec:results_MN_potential} before we move on to the more realistic local GT+ chiral potentials at N$^2$LO in Sec.~\ref{sec:results_chiral_potential}.  

\subsection{Minnesota potential}
\label{sec:results_MN_potential}

\begin{figure}[tbh!]
    \centering
    \includegraphics[width=\columnwidth]{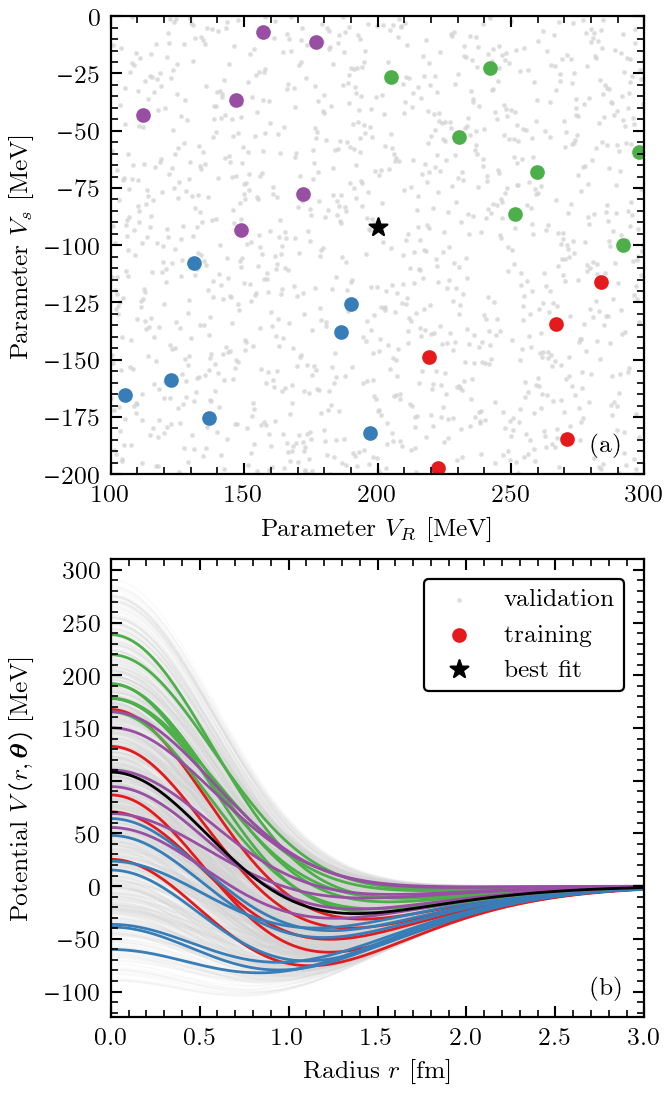}
    \caption{Illustration of the Minnesota (MN) potential. 
    Panel~(a) shows the potential's two-dimensional parameter space $(V_R,V_s)$. 
    The potential's other two parameters are kept at their respective best-fit values.
    The training set (large dots) contains $25$ points, whereas the denser validation set (smaller dots) contains $1250$ points.
    For illustration, the parameter space is arbitrarily divided into four quadrants, where the training points in those quadrants are colored differently. 
    The black star in panel~(a) indicates the best-fit value.
    Panel~(b) depicts the potential's dependence on the radial coordinate, $V(r; \paramVec)$, corresponding to the parameters depicted in the top panel. 
    Both panels use the same color coding.
    See the main text for details.}
    \label{fig:minnesota-comparison-space}
\end{figure}

Following Refs.~\cite{Furnstahl:2020abp,Melendez:2021lyq}, we consider the Minnesota potential in the \oneSzero ($\ell =0$) channel~\cite{THOMPSON197753}:
\begin{equation} \label{eq:minnesota}
    V(r;\paramVec) = V_{R} \, e^{-\kappa_R r^2} + V_{s} \, e^{-\kappa_s r^2} \,,
\end{equation}
which was calibrated in Ref.~\cite{THOMPSON197753} to reproduce experimentally extracted effective range parameters, resulting in the best-fit values for the four model parameters: $V_{R} = 200 \MeV$, $V_{s} = -91.85\MeV$, $\kappa_R = 1.487\fmisq$, and $\kappa_s = 0.465\fmisq$.
Because the Minnesota potential~\eqref{eq:minnesota} is affine only in $V_{R}$ and $V_{s}$, we vary only those parameters, i.e., $\paramVec = \{ 1, V_{R}, V_{s}\}$,%
\footnote{See Sec.~\ref{subsec:rse}, and particularly Eq.~\eqref{eq:affine_potential}, where we discuss the auxiliary dimension (i.e., the first component of $\paramVec$) associated with the parameter-independent part of the affine decomposition~\eqref{eq:affine_potential}.} 
and keep $\kappa_R$ and $\kappa_s$ constant at their corresponding best-fit values.

Figure~\ref{fig:minnesota-comparison-space} shows the potential's remaining two-dimensional parameter space~[Fig.~\ref{fig:minnesota-comparison-space}~(a)] and the associated potentials [Fig.~\ref{fig:minnesota-comparison-space}~(b)] as a function of the radial coordinate, $V(r; \paramVec)$.
The black star in Fig.~\ref{fig:minnesota-comparison-space}~(a) indicates the potential's best-fit values.
In the following, we consider the training set with $25$ points depicted by the large dots in Fig.~\ref{fig:minnesota-comparison-space}~(a) and the denser validation set with $1250$ additional points depicted by the smaller dots.
These sets were obtained independently by LHS in the parameter ranges shown in Fig.~\ref{fig:minnesota-comparison-space}: $V_{R} = [100, 300] \MeV$ and $V_{s} = [-200, 0] \MeV$, which were chosen as in Ref.~\cite{Furnstahl:2020abp} to encompass the best-fit value.
To illustrate that the behavior of the potential changes significantly in this parameter space, as depicted in Fig.~\ref{fig:minnesota-comparison-space}~(b), including mainly repulsive (green lines) and attractive (blue lines) potentials, we arbitrarily divide the potential's parameter space into four quadrants and color the training points in those quadrants differently.
Both panels use the same color coding.

\begin{figure*}[tb]
    \centering
    \includegraphics[width=\textwidth]{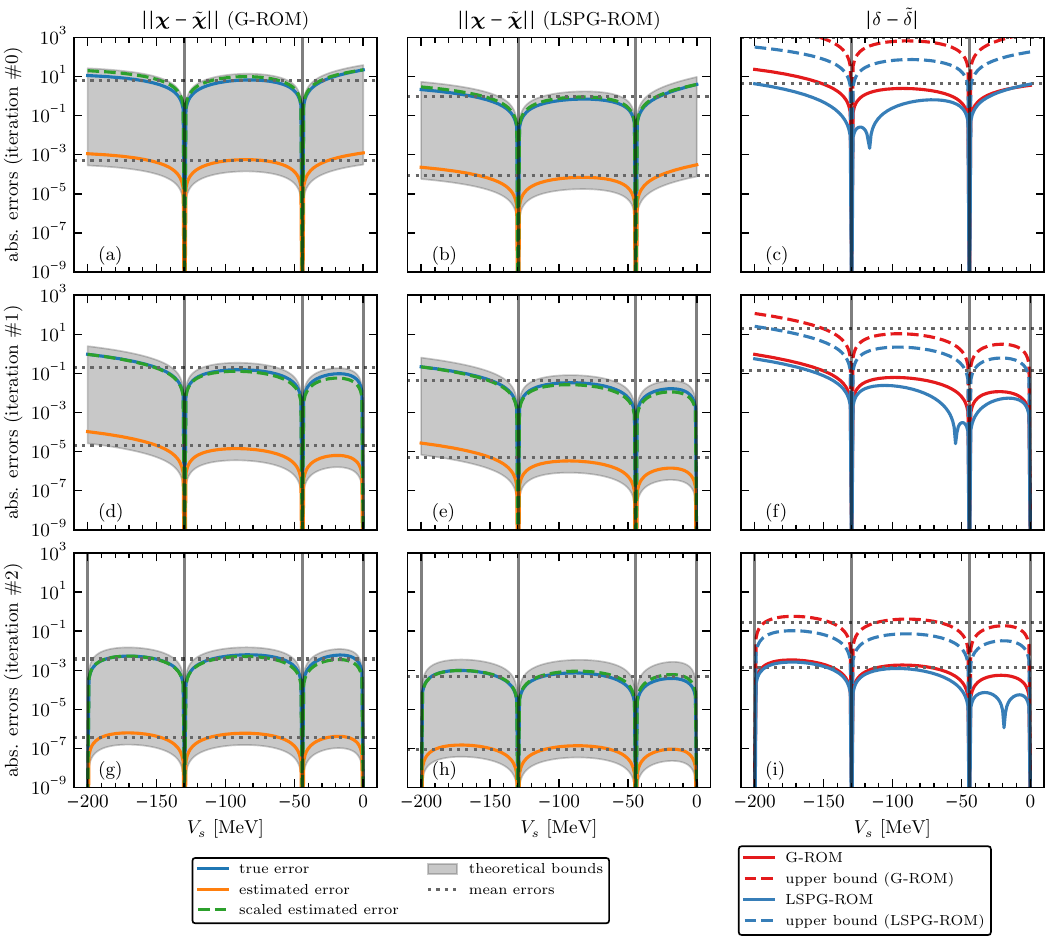}
    \caption{%
    Estimated and true absolute errors obtained by the G-ROM (left column) and LSPG-ROM (center column) emulators for the Minnesota potential~\eqref{eq:minnesota} with a single parameter varied, $V_s$, at fixed $V_R$, $\kappa_R$, and $\kappa_S$. 
    The left and center columns show the absolute errors in the (unmatched) wave functions; the right column shows the corresponding absolute errors (i.e., their theoretical upper bounds) in the scattering phase shifts.
    We choose $E = 50 \MeV$ for this demonstration.
    The top row shows the emulators' initial configurations.
    Each successive row is another iteration of the greedy algorithm, in which an additional snapshot is calculated at the location of the previous iteration's largest estimated error. 
    Both emulators scan the same parameter spaces, start with the same initial snapshots, and run for two iterations. 
    The number sign ``\#'' in the $y$-axis labels specifies the number of iterations conducted by the greedy algorithm.
    See the main text for details.}
    \label{fig:G-ROM_LSPG_comparison}
\end{figure*}

Figure~\ref{fig:G-ROM_LSPG_comparison} demonstrates the key components of the greedy algorithm: error estimation and iterative placement of snapshots where the estimated error is the largest, thereby reducing the local (and global) emulator error for both the wave functions and scattering phase shifts.
For demonstration, we vary only $V_{s}$ while keeping $V_{R}$ fixed at its best-fit value and choose $E = 50 \MeV$.
This variation is conducted along a vertical line through the black star in the top panel of Fig.~\ref{fig:minnesota-comparison-space}.
We randomly place two initial snapshots via LHS and let the greedy algorithm iterate twice.
Given these initial snapshots, the emulator has to both interpolate and extrapolate in the parameter range to run the greedy algorithm for training.

From the top to the bottom, the rows in Fig.~\ref{fig:G-ROM_LSPG_comparison} correspond to absolute error in the (unmatched) wavefunctions given the emulators' current bases after each successive iteration, starting with the initial state shown in the top row [i.e., Fig.~\ref{fig:G-ROM_LSPG_comparison}~(a)--\ref{fig:G-ROM_LSPG_comparison}(c)]. 
The gray vertical lines in all panels indicate the snapshot locations used for training, which coincide with the locations of vanishing emulator errors.
The left and center columns show the estimated (orange lines) and true (blue lines) errors of the scattered wave functions across the varied parameter space for the G-ROM and LSPG-ROM, respectively.
The right column ([i.e., \ref{fig:G-ROM_LSPG_comparison}~(c), \ref{fig:G-ROM_LSPG_comparison}~(f), and \ref{fig:G-ROM_LSPG_comparison}~(i)] shows the associated errors in the scattering phase shifts for the two emulators, and the dashed lines depict the upper error bounds derived in Eqs.~\eqref{eq:norm_error_delta} and~\eqref{eq:lsq_sol_a_b_error}.
These error bounds turn out to be very conservative.
Note that the two emulators use the same initial snapshots and scan in the same parameter range for the maximum (estimated) error.
Both emulators exhibit comparable errors across the parameter range, making them choose the same snapshots iteratively.
This behavior may not always be so.

As one can see, 
the true and estimated errors in the left and center columns of Fig.~\ref{fig:G-ROM_LSPG_comparison}
are approximately proportional to each other, allowing for the simple calibration of the estimated error discussed in Sec.~\ref{sec:iterative_improvement}.
This calibration rescales the estimated error by their ratio, which is evaluated at the parameter location of the largest estimated error and depicted by the dashed green lines. 
The dotted horizontal lines depict the mean errors across the parameter range shown for the estimated (lower lines) and rescaled estimated error (upper lines).
The gray uncertainty band depicts the theoretical bounds~\eqref{eq:error_bounds} on the emulator errors for the wave functions, which we computed here only for illustration purposes, as it requires a relatively expensive SVD of the FOM matrix.
We find that the true error of the emulated wave functions is close to the upper theoretical bound in Eq.~\eqref{eq:error_bounds}, indicating that it may indeed serve as both a rigorous and not-too-conservative error estimate if the smallest singular value could be efficiently estimated.
After two iterations, both emulators obtain a mean absolute error of about $10^{-3}$ or better for the phase shifts across the considered parameter space.

Figure~\ref{fig:G-ROM_LSPG_comparison} also shows that the greedy algorithm iteratively chooses the boundaries of the parameter space as snapshot locations. 
The emulator has to extrapolate in those regions (before adding these snapshots to the emulator basis), so the errors can be expected to be largest there.
One might be tempted to always place snapshots at all or some randomly chosen boundaries to minimize emulator errors.
In high-dimensional spaces, this strategy is computationally challenging at best due to the curse of dimensionality.
However, by minimizing the estimated emulator error, the greedy algorithm provides the means to choose a requested number of snapshot locations, including the boundaries of the parameter space, that are locally optimal.
This feature may be crucial for applications to high-dimensional problems.

\begin{figure}[tb]
    \centering
    \includegraphics[width=\columnwidth]{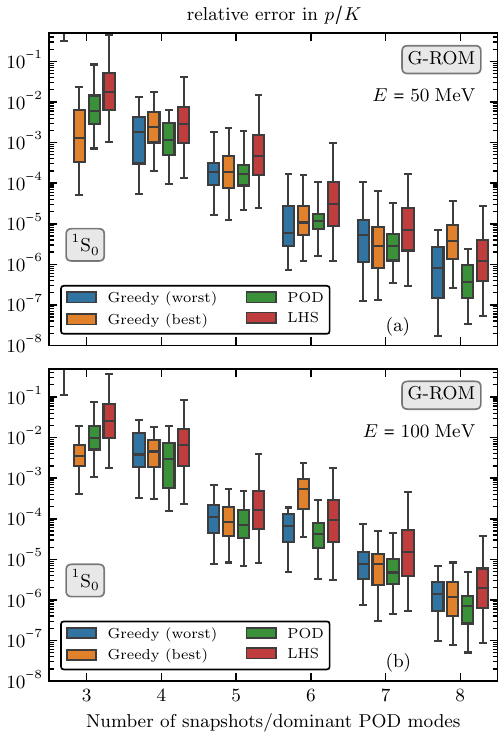}
    \caption{%
    Emulator convergence analysis and comparison at $E=50 \MeV$ in panel~(a) and $100 \MeV$ in panel~(b) for the Minnesota potential~\eqref{eq:minnesota} in the \oneSzero channel. 
    The emulator basis constructed using the greedy algorithm is compared with the LHS and POD approaches. 
    The three emulators have access to the same 25 training points in the two-dimensional parameter space depicted in Fig.~\ref{fig:minnesota-comparison-space}; but only the POD-based emulator is informed by all training points, whereas the greedy and LHS emulators use a smaller subset of these training points (as indicated by the $x$ axis). 
    The accuracy of these emulators is assessed using the validation set of 1250 points depicted in Fig.~\ref{fig:minnesota-comparison-space}.
    Each box shows the range between the first and third quartile of the samples across the validation points, with the median indicated by the horizontal line inside the box.
    The whiskers extend the quartiles to the 5th and 95th percentiles, respectively.
    See the main text for details.}
    \label{fig:minnesota-emulator-comparison-grom}
\end{figure}

\begin{figure}[tb]
    \centering
    \includegraphics[width=\columnwidth]{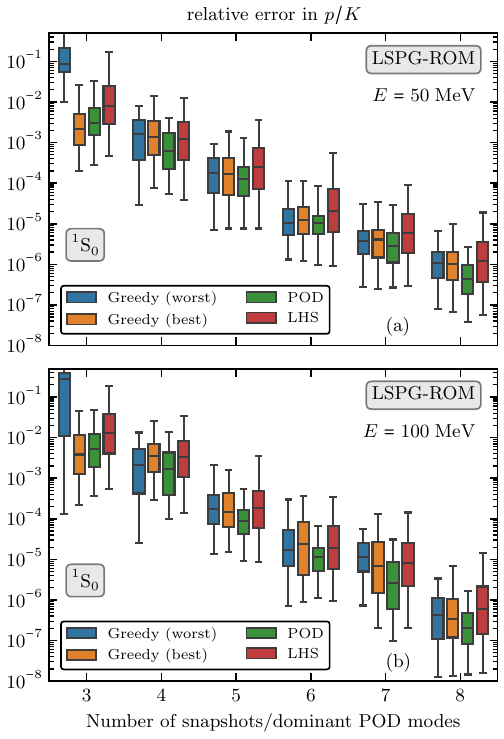}
    \caption{%
    Same as Fig.~\ref{fig:minnesota-emulator-comparison-grom} but for the LSPG-ROM.%
    }
    \label{fig:minnesota-emulator-comparison-lspg}
\end{figure}

Next, we consider the Minnesota potential's two-dimensional parameter space depicted in Fig.~\ref{fig:minnesota-comparison-space}, varying $V_{R}$ and $V_{s}$ simultaneously.
To assess the greedy algorithm's convergence, we study the exact emulator errors as a function of the number of snapshots in the emulator basis and contrast it with those obtained with two alternative approaches in Fig.~\ref{fig:minnesota-emulator-comparison-grom}: 
the POD approach discussed in Sec.~\ref{sec:pod} and randomly choosing a subset of the 25 training points (LHS).
It is important to note that only the POD approach performs and uses FOM calculations at all 25 training points. 
In contrast, the other two approaches only perform FOM calculations for a subset of these 25 training points.

Figures~\ref{fig:minnesota-emulator-comparison-grom} (G-ROM) and~\ref{fig:minnesota-emulator-comparison-lspg} (LSPG-ROM) show the resulting convergence pattern of the symmetric relative error defined by
\begin{equation} \label{eq:sym_rel_err}
    \epsilon(\tilde{x}) = 2\frac{|x - \tilde{x}|}{|x| + |\tilde{x}|} \,
\end{equation}
across the validation points depicted in Fig.~\ref{fig:minnesota-comparison-space}. 
Here, the emulator approximation for the FOM calculation of the variable $x$ is denoted by $\tilde{x}$.
Following Ref.~\cite{Furnstahl:2020abp}, we choose $x = p / K$ for this comparison. 
The two panels in Figs.~\ref{fig:minnesota-emulator-comparison-grom} and~\ref{fig:minnesota-emulator-comparison-lspg} show the so-called box plots of these errors for two representative energies: $E=50 \MeV$ in Figs.~\ref{fig:minnesota-emulator-comparison-grom}~(a) and \ref{fig:minnesota-emulator-comparison-lspg}~(a) and $E=100 \MeV$ in Figs.~\ref{fig:minnesota-emulator-comparison-grom}~(b) and \ref{fig:minnesota-emulator-comparison-lspg}~(b).
Each box depicts the range between the first and third quartile of the samples across the validation points (i.e., the inter-quartile range), with the median indicated by the horizontal line inside the box.
The whiskers extend the quartiles to the 5th and 95th percentiles, respectively.
We observe systematic convergence patterns for both the G-ROM and LSPG-ROM: 
the median of the symmetric relative error~\eqref{eq:sym_rel_err} decreases exponentially with increasing size of the emulator basis.

To start the greedy algorithm, we must choose the number of initial snapshots and their locations in the parameter space. 
We set $n_b = 3$, which gives us ${25 \choose 3} =2300$ possible combinations to choose these initial snapshots given the 25 training points depicted in Fig.~\ref{fig:minnesota-comparison-space}.
We construct all ${25 \choose 3}$ emulators, orthornormalize their snapshot bases (without compression), and benchmark their accuracies as measured by the median of the symmetric relative error~\eqref{eq:sym_rel_err} across the set of validation points.
In the following, we consider only two of these emulators: 
the ones with the smallest and the largest median errors.
We refer to them as ``Greedy (best)'' and ``Greedy (worst),'' respectively.
They are depicted by the blue and orange boxes in Figs.~\ref{fig:minnesota-emulator-comparison-grom} and~\ref{fig:minnesota-emulator-comparison-lspg}.
Although the ``Greedy (worst)'' emulator is significantly less accurate for the given initial setting (i.e., before the first greedy iteration), it consistently recovers after only one iteration, leading to similar accuracies as the ``Greedy (best)'' emulator throughout the iteration process, as measured by the inter-quartile ranges.
This overall behavior emphasizes that, even though the emulator accuracy is sensitive to the chosen snapshot basis, the proposed greedy algorithm can efficiently identify and remedy poor choices of (initial) snapshots.

The two greedy emulators are contrasted with a POD emulator, which is trained on all 25 training points and depicted by the green boxes in Figs.~\ref{fig:minnesota-emulator-comparison-grom} and~\ref{fig:minnesota-emulator-comparison-lspg}.
Its emulator basis is then truncated to the dominant POD modes indicated by the $x$-axis.
This POD compression is aggressive if only a few POD modes are considered:
one would need to include 13 POD modes to achieve a machine precision-level truncation of the singular values.
Note that the $x$-axis in Figs.~\ref{fig:minnesota-emulator-comparison-grom} and~\ref{fig:minnesota-emulator-comparison-lspg} ranges only up to eight dominant POD modes.
We also contrast the greedy emulators with entirely random subsets of the training points in Fig.~\ref{fig:minnesota-comparison-space}.
There are ${25 \choose n_b}$ possible combinations to build these emulators. 
We sample only up to 100 of them to benchmark the accuracy of this random approach, referred to as ``LHS.''
The bases of the four emulators have the number of basis vectors $n_b$ indicated by the $x$-axis in Figs.~\ref{fig:minnesota-emulator-comparison-grom} and~\ref{fig:minnesota-emulator-comparison-lspg}. 

As shown in Figs.~\ref{fig:minnesota-emulator-comparison-grom} and~\ref{fig:minnesota-emulator-comparison-lspg}, the basis constructed using the greedy algorithms after at least one iteration predict inter-quartile ranges comparable to the POD and LHS approaches across the validation points. 
However, the na{\"i}ve LHS approach exhibits the largest variation in the errors, which may lead to overall unreliable results, especially since this approach does not estimate errors, and space-filling sampling may not be feasible for higher-dimensional parameter spaces. 
In other words, the constructed emulators are similar in accuracy but not necessarily in precision.
The greedy and POD approaches consistently produce accurate and precise predictions for $n_b >3$.
This finding might be unsurprising for the POD approach as it has access to the most information and uses the most significant emulator basis vectors (i.e., POD modes) as measured by their associated singular values. 
However, the POD approach's drawback is that it requires far more high-fidelity calculations than the other emulators and does not estimate errors, unlike the greedy emulators.
The greedy emulators achieve high accuracies similar to the POD approach while using far fewer high-fidelity calculations to construct their bases and estimate errors.
Overall, we find similar convergence patterns for the G-ROM and LSPG-ROM. 
However, the LSPG-ROM has more consistent symmetric relative errors (i.e., fewer fluctuations) than the G-ROM, which may be related to the presence of Kohn anomalies studied in Sec.~\ref{sec:results_chiral_potential}.

For completeness, we note that one could use the Kohn variational principle (KVP) to further improve the emulated $K$-matrix element based on our emulated wave functions~\cite{Furnstahl:2020abp,Garcia:2023slj}.
However, the KVP does not treat the $K$-matrix element and the associated wave function on an equal footing:
it only provides an improved estimate for the $K$-matrix element, which is second order (as opposed to linear) in the error, not for the wave function.
It also does not provide error estimates in contrast to our approach.
Even without this Kohn correction, our emulated $K$-matrix elements are similarly, although slightly less, accurate than the results shown in Fig.~3 in Ref.~\cite{Furnstahl:2020abp}.
Hence, we do not apply the Kohn correction here.

\subsection{Local GT+ chiral potential}
\label{sec:results_chiral_potential}

Next, we apply the greedy emulator to the more realistic GT+ chiral potential at N$^2$LO~\cite{Gezerlis:2014zia} with the regulator cutoff $R_0 = 1.0 \fm$ and spectral function cutoff $\tilde{\Lambda} = 1000 \MeV$. 
These potentials are fit to NN scattering data and are commonly used in microscopic calculations of finite nuclei and infinite matter.
As discussed in detail in Appendix~A in Ref.~\cite{Gezerlis:2014zia}, the N$^2$LO interaction in the $np$ channel generally depends on nine low-energy constants (LECs), which form $\paramVec = \{ 1, C_S, C_T, C_1, C_2, \ldots, C_7\}$.\footnote{Recall that the first dimension of the parameter vector corresponds to the parameter-independent part of the interactions, which contains here all of the pion-exchange physics. See Eq.~\eqref{eq:affine_potential}. The pion-nucleon couplings $c_1$, $c_3$, and $c_4$ are kept at their best-fit values~\cite{Gezerlis:2014zia}.} 
Following Ref.~\cite{Drischler:2021qoy}, we vary all of these nine parameters, although not all contribute to every partial-wave channel. 
Typically, only a few linear combinations of these LECs contribute to any partial-wave channel, which could be used for dimensionality reduction.
For example, two linear combinations contribute to the \oneSzero channel and three to the \threePzero channel at N$^2$LO. 
More details can be found in Appendix~A in Ref.~\cite{Gezerlis:2014zia}.
The developers provided their GT+ potentials as a \texttt{C\textsuperscript{++}} code, which we modified to output the affine components $\mathcal{V}_a^{(\ell)}(r)$ in Eq.~\eqref{eq:affine_potential} and the best-fit values of the LECs.
More details on this family of chiral interactions can be found in Ref.~\cite{Gezerlis:2014zia}. 

\begin{figure}[tb]
    \centering
    \includegraphics[width=\columnwidth]{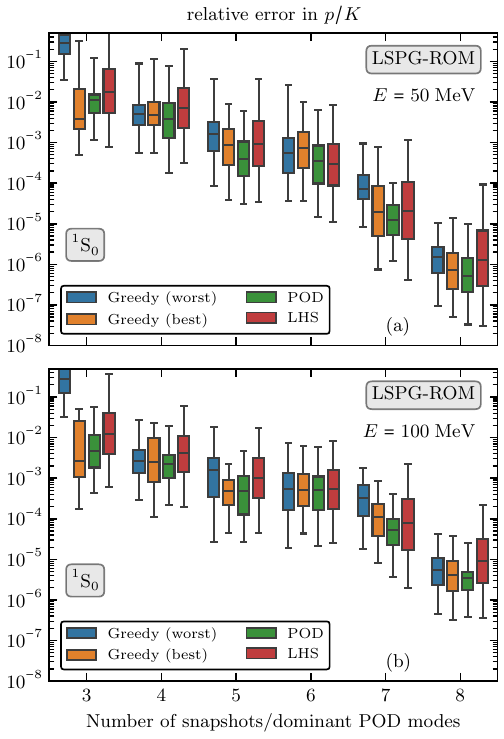}
    \caption{Same as Fig.~\ref{fig:minnesota-emulator-comparison-lspg} (LSPG-ROM) but for the \nTwoLO GT+ potential with $R_0 = 1.0 \fm$ and $\tilde{\Lambda} = 1000 \MeV$ in the \oneSzero channel. 
    The training set consists of $200$ and the validation set of $10^4$ random points, drawn using LHS in the nine-dimensional $\pm 50\%$ region around the LECs' best-fit values in their respective units.
    Not all of these LECs contribute to this channel.
    See the main text for details.}
    \label{fig:chiral-1S0-emulator-comparison-lspg}
\end{figure}

\begin{figure}[tb]
    \centering
    \includegraphics[width=\columnwidth]{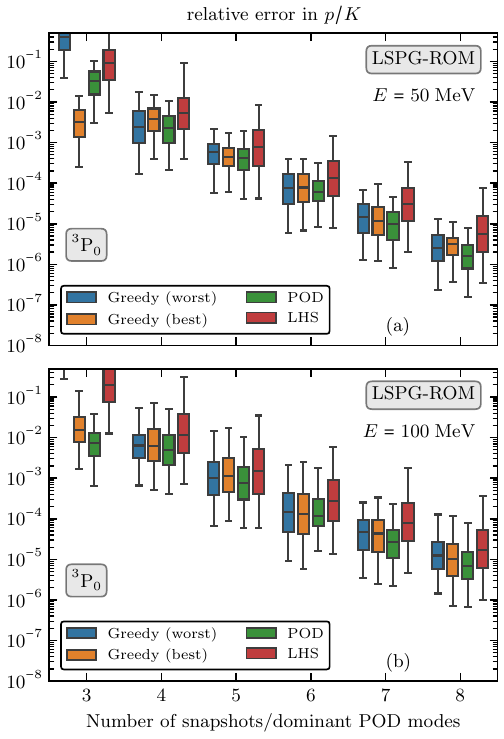}
    \caption{Same as Fig.~\ref{fig:chiral-1S0-emulator-comparison-lspg} (LSPG-ROM, \nTwoLO GT+ potential) but in the \threePzero ($\ell = 1$) channel.}
    \label{fig:chiral-3P0-emulator-comparison-lspg}
\end{figure}

As we did for the Minnesota potential in Sec.~\ref{sec:results_MN_potential}, we study the convergence of the greedy emulator in comparison with the POD and na{\"i}ve LHS approach to snapshot selection. 
Figures~\ref{fig:chiral-1S0-emulator-comparison-lspg} and~\ref{fig:chiral-3P0-emulator-comparison-lspg} show our results for this comparison in the \oneSzero and \threePzero channels, respectively, similar to Fig.~\ref{fig:minnesota-emulator-comparison-lspg} for the Minnesota potential and LSPG-ROM.
Given the significantly higher-dimensional parameter space of the chiral interaction, we increase the number of training points to $200$ and the number of validation points to $10^4$ points, both obtained via LHS in the $\pm 50\%$ region around the LECs' best-fit values in their respective units used in Table~I of Ref.~\cite{Gezerlis:2014zia}.
From the ${200 \choose 3} = 1\,313\,400$ possible initial snapshot locations, we randomly select 100 to estimate the best and the worst configurations to initialize the greedy algorithm. 
Likewise, we randomly sample only 400 of the ${200 \choose n_b}$ combinations for initializing the LHS approach because of the large number of emulators possible to construct.
We focus here on the LSPG-ROM and these two channels as representative test cases since we found qualitatively similar convergence patterns for the G-ROM and higher partial-wave channels.
More test cases can be found in the companion GitHub repository~\cite{BUQEYEsoftware}, which contains all the source codes needed to reproduce and extend our analysis.

Overall, we find convergence patterns independent of the partial-wave channels and similar to those for the \oneSzero Minnesota potential in Sec.~\ref{sec:results_MN_potential}.
In particular, we find again that the greedy algorithm can remedy a poor initial choice for the emulator basis (``Greedy (worst)'') and obtains high accuracies comparable to the more informed POD approach but at a lower computational cost. 
Both approaches have higher precision and accuracy than the na{\"i}ve LHS approach.

\begin{figure}[tb]
    \centering
 \includegraphics[width=0.98\linewidth]{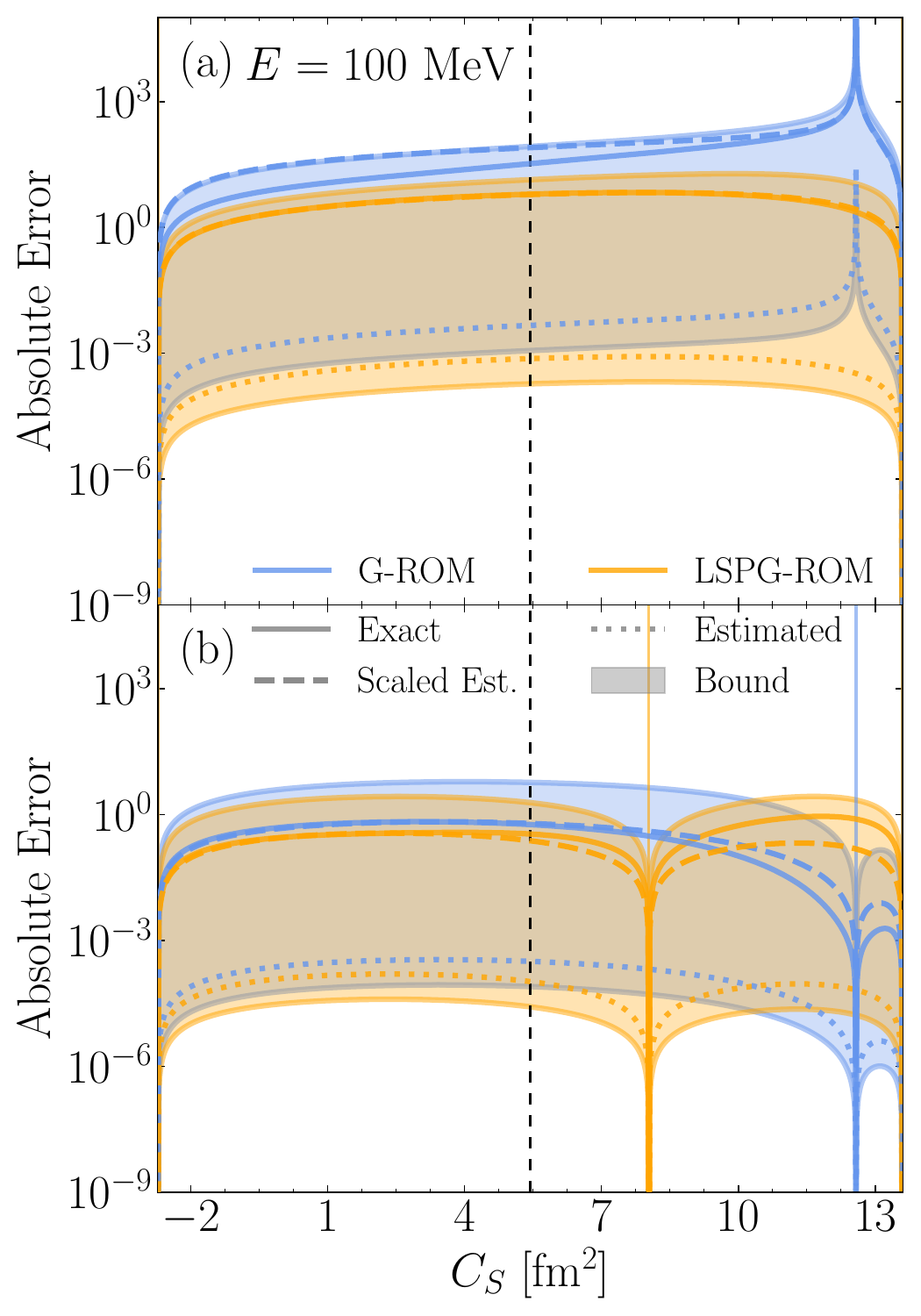}
    \caption{%
    Absolute emulator errors in the scattered wave functions showing the (a)~visualization with 2 snapshots and~(b) elimination with 3 snapshots of a Kohn anomaly when using the G-ROM emulator (blue bands) for the \nTwoLO GT+ chiral potential in the \oneSzero channel at $E = 100\MeV$. 
    The errors are shown as a function of the LO LEC $C_S$.
    The Kohn anomaly
    at $C_S \approx 12.6 \fmsq$  dictates the location of the next snapshot of the G-ROM's greedy algorithm, removing it in the process. 
    We have seen no such anomalies for the LSPG-ROM emulator (orange bands),
    which does not have its snapshot basis influenced.
    The vertical dashed black line in the center of the two panels shows the best-fit value of $C_S = 5.4385 \fmsq$ obtained in Ref.~\cite{Gezerlis:2014zia}, around which the parameter space was created.%
    }
    \label{fig:anomaly-visualization}
\end{figure}

Next, we explore the Kohn anomalies discussed in Sec~\ref{sec:lspg}, where we found analytical evidence that the LSPG-ROM is less susceptible to those spurious singularities than the G-ROM.
Figure~\ref{fig:anomaly-visualization} shows the absolute errors in the scattered wave functions obtained with the G-ROM and LSPG-ROM across the one-dimensional parameter space of the leading order (LO) LEC $C_S$ at two consecutive steps in the greedy iteration.
The two initial snapshots in Fig.~\ref{fig:anomaly-visualization}~(a) are placed at $C_S = -2.7 $ and $13.6 \fmsq$; i.e., at the boundaries of the $C_S$ range shown. 
Each ROM emulates at the same 900 training points in $C_S \in [-2.7, 13.6] \fmsq$.
For illustration, the $C_S$ parameter range depicted in Fig.~\ref{fig:anomaly-visualization} was determined by a $150\%$ variation around the best-fit value $C_S = 5.4385 \fmsq$ (dashed vertical line), as determined in Ref.~\cite{Gezerlis:2014zia}.

With this set of snapshot parameters, the G-ROM exhibits a Kohn anomaly at $C_S \approx 12.6 \fmsq$, while the LSPG-ROM does not. 
We generally observed no anomaly with the LSPG-ROM in our extensive parameter sweeps.
The location of this anomaly has maximum error, so the greedy algorithm determines that this snapshot will be added to the G-ROM emulator basis, thereby removing the Kohn anomaly.
This removal process can be seen in Fig.~\ref{fig:anomaly-visualization}~(b). 
In contrast, the greedy algorithm for the LSPG-ROM can place a snapshot at a different parameter value to improve the emulator basis without needing to remove an anomaly.
This feature may give it an advantage over the G-ROM in terms of efficacy, especially in high-dimensional parameter spaces.
Overall, we find that the greedy algorithm can detect and remove Kohn anomalies similar to the emulator-mixing method developed in Ref.~\cite{Drischler:2021qoy} but comes with error estimation.

Finally, we discuss the computational speed-up factors of our emulators. 
Once the emulators are trained, we find implementation-dependent speed-up factors of about $2-6$ in the emulator's online stage compared to the FOM solver.
At first glance, these speed-up factors seem lower than those of scattering emulators already published.
However, we emphasize that our FOM solver implementing the matrix Numerov method is highly computationally efficient due to the LAPACK implementation and the offline-online decomposition (for potentials with affine parameter dependences) even at the FOM level, as discussed in Sec.~\ref{sec:matrix_numerov}.
Furthermore, we focus here on algorithmic development, not the most efficient computational implementation. 
We emphasize that speed-up factors are both implementation and hardware-dependent, which one should consider when interpreting and comparing them.

Disregarding computational overhead, the greedy algorithm's speed-up in the emulator's offline stage compared to the POD approach is because the greedy algorithm (ideally) requires fewer FOM solutions to train the emulator. 
For example, if the POD approach determines only five dominant POD modes based on 200 FOM calculations and the greedy algorithm obtains similar accuracy with just five FOM calculations, the theoretical speed-up factor is 40. 
This scenario is depicted in Figs.~\ref{fig:chiral-1S0-emulator-comparison-lspg} and~\ref{fig:chiral-3P0-emulator-comparison-lspg} for the GT+ chiral potential.
The mentioned speed-up factor is only an upper bound due to the computational overheads of both the POD approach and the greedy algorithm.
On the POD side, the overhead is mainly due to the need to perform an SVD of the snapshot basis with all 200 FOM solutions, which is an expensive numerical operation; 
on the greedy algorithm side, the overhead is mainly due to the need to emulate the scattering solutions on the discretized parameter space and find the location of the maximum estimated error; hence, the overhead scales with the speed-up factor of the emulator's online stage.
In conclusion, we expect the greedy algorithm to shine in cases where FOM calculations are computationally very expensive to prohibitively slow, such as coupled cluster calculations of nuclear structure~\cite{PhysRevLett.123.252501} and three-body scattering~\cite{Zhang:2021jmi}. 
On the other hand, in the case of NN scattering, the scattering equations can already be accurately and efficiently solved, which is an ideal test case for our algorithmic developments.

\section{Summary and Outlook}
\label{sec:summary_outlook}

In this paper, we used a prototypical test case for emulators, NN scattering in coordinate space, to guide the development and implementation of two active learning emulators based on the (Petrov-)Galerkin projection methods: the G-ROM and LSPG-ROM.
Both emulators are implemented and tested using the matrix Numerov method, a reformulation of the Numerov recurrence relation as a system of coupled linear equations, for solving the homogeneous or inhomogeneous RSE.
We chose this high-fidelity ODE solver because of its high accuracy and popularity in the nuclear physics community. 
However, our emulators are widely applicable to any ODE solver and physics problem that can be expressed as linear systems. 

The emulators have error estimators, enabling efficient emulator basis selection via the developed greedy algorithm.
This algorithm iteratively places snapshots in the model parameter space where the estimated emulator error is the largest, thereby systematically improving the emulator's accuracy.
Assuming an affine parameter dependence of the nuclear interactions, we developed efficient offline-online decompositions for the two emulators and their error estimators, which is key to obtaining high computational speed-up factors.
These developments set the groundwork for further applications to NN and three-nucleon chiral interactions~\cite{Zhang:2021jmi,Wesolowski:2021cni} and optical models~\cite{Odell:2023cun} currently in progress.

Following the general philosophy of this work as providing a prototype for emulation, we discussed the matrix Numerov method (see Sec.~\ref{sec:fom}), derived the emulator equations (see Sec.~\ref{sec:rom}), and detailed the greedy algorithm as an alternative to the POD approach (see Sec.~\ref{sec:snapshot_selection}) with sufficient detail to enable both the reproduction and extension of our results to other problems.
After these comprehensive discussions, we first illustrated the greedy algorithm in the controlled case of the simple Minnesota potential~\cite{THOMPSON197753}. 
We then applied the approach to the more realistic chiral NN interactions at N$^2$LO, called GT+~\cite{Gezerlis:2014zia}, which are commonly used in modern ab initio calculations of finite nuclei and infinite matter, and benchmarked the resulting emulators against the POD approach to snapshot selection.

In general, as expected, we found that the POD approach consistently obtains high accuracies because it contains the most information on high-fidelity solutions across the parameter space through space-filling sampling.
However, this high information content comes at the computational expense of performing many high-fidelity calculations in the emulator's offline stage, which can be prohibitively slow and even unnecessary if high POD compression rates are obtained. 
On the other hand, the greedy algorithm achieves similar accuracies (for equal emulator basis sizes) while requiring significantly fewer high-fidelity calculations in the offline stage due to its active learning approach to snapshot selection.
It also inherently comes with an error estimator, unlike the POD approach.
Hence, the developed greedy algorithm may facilitate the training of fast \& accurate emulators that would otherwise be computationally expensive or even unaffordable.

Model-driven scattering emulators are susceptible to spurious singularities known as Kohn (or Schwartz) anomalies. 
These anomalies occur when the emulator equations are singular or near singular.
We thoroughly searched for the presence of these anomalies in the G-ROM and LSPG-ROM, but found only anomalies in the G-ROM, which is due to the special analytic structure of the LSPG-ROM that mitigates anomalies.
We also demonstrated that the greedy algorithm can detect and mitigate Kohn anomalies by placing snapshots in the parameter space where anomalies occur (see Fig.~\ref{fig:anomaly-visualization}). 
Efficient Kohn anomaly detection and mitigation are important when using emulators in practice.

Developing efficient methods for estimating the smallest singular value of the non-Hermitian FOM matrix is crucial for deriving conservative error estimators.
We observed that the estimated emulator error is approximately proportional to the true error. 
Under this assumption, which should be validated in each application, one can calibrate the error estimator via rescaling at the expense of an additional high-fidelity calculation. 
Estimating the smallest singular value efficiently would allow us to use the derived theoretical upper bound on the estimated error as a conservative and more rigorous error estimate without additional high-fidelity calculations. 
Methods for estimating extremal singular values exist, including the studied SCM~\cite{HUYNH2007473}.
However, we found that the SCM is too expensive in our particular case because the lower bound for the smallest singular value is significantly underestimated (while the upper bound is remarkably close to the actual smallest singular value). 

The offline-online decompositions of our emulators require affine parameter dependences of the underlying potentials.
While the NN contact LECs of chiral nuclear interactions, which were the main focus here, meet this requirement, optical models are precluded. 
However, the empirical interpolation method (EIM) has already been demonstrated~\cite{Odell:2023cun} to be an efficient hyperreduction method to render the parameter dependences of optical potentials approximatively affine. 
Combined with the EIM, our emulators, therefore, are also applicable to potentials with non-affine parameter dependences.
Explorations of optical models will benefit from the fact that the derived emulator equations are valid for both real- and complex-valued potentials.

This work can be applied and extended in various ways. 
The next steps include extending the scattering emulator to coupled partial-wave channels, momentum space interactions, and the long-range Coulomb interaction, e.g., using the Vincent-Phatak method~\cite{Melendez:2021lyq,Vincent:1974zz}.
These extensions are needed for calculating cross-sections and spin observables and thus calibrating nuclear interactions and optical models with quantified uncertainties directly to experimental data. 
One way to achieve these extensions would be to apply our machinery to the Lippmann-Schwinger integral equation (in momentum space), which would give direct access to momentum-space interactions and coupled channel scattering.

Another step for this line of research is to investigate three-nucleon (and higher-body) scattering, where runtimes of high-fidelity solvers are no longer computationally tractable, not even for emulator training via the POD approach.
The greedy algorithm developed here combined with an efficient emulator for three-body scattering, such as the one proposed in Ref.~\cite{Zhang:2021jmi} based on the Kohn variational principle, would play a crucial role in the uncertainty quantification of chiral three-body (and higher-body) forces to constrain next-generation chiral interactions and study modified EFT power counting schemes (e.g., see Refs.~\cite{Yang:2020pgi,Drischler:2022yfb,Cirigliano:2024ocg,Sekiguchi:2024rka}).
The DOE STREAMLINE collaboration is working on these applications of active learning emulators for nuclear scattering.

\begin{acknowledgments}

We thank S.~Rave and our DOE STREAMLINE collaborators, especially Ch.~Elster, A.~Giri, and J.~Kim, for fruitful discussions, and I.~Tews for sharing the source code that evaluates the GT+ chiral potentials~\cite{Gezerlis:2014zia}.
We are also grateful to the developers of the Model Order Reduction with Python package \texttt{pyMOR}~\cite{pymor} for many insights into active learning algorithms.
We also acknowledge fruitful discussions with B.~McClung and A.~C.~Semposki.
This material is based upon work supported by the U.S.~Department of Energy, Office of Science, Office of Nuclear Physics, under the FRIB Theory Alliance Award No. DE-SC0013617, under the STREAMLINE collaboration Awards DE-SC0024233 (Ohio University) and DE-SC0024509 (Ohio State University), and by the National Science Foundation under Award No. PHY-2209442.
RJF also acknowledges support from the ExtreMe Matter Institute EMMI
at the GSI Helmholtzzentrum für Schwerionenforschung GmbH, Darmstadt, Germany.

\end{acknowledgments}

\appendix
\section{Numerov method for initial value problems}
\label{app:init_gonzalez}

This appendix discusses, for completeness, the Numerov method for initial value problems in which values for $(y_0, y'_0)$ are imposed on the solution of the ODE~\eqref{eqn:formal-numerov}. 
The local accuracy of estimating $y_1$ may be crucial for obtaining globally accurate solutions.

Here, we present the work by Gonz{\'a}lez \& Thompson~\cite{Gonzalez}, who considered the series expansion of the ODE's solution about $r_0 = 0$:
\begin{equation} \label{eq:gonzalez_series}
    y_1 = y_0 + h y'_0 
    + \frac{h^2}{2!} f_0 
    + \frac{h^3}{3!} f'_0
    + \frac{h^4}{4!} f''_0
    + O(h^5) \,,
\end{equation}
with the short-hand notation $f_n = f(r_n; y_n)$ and similarly for its spatial derivatives.
The initial condition and step size determine the first two terms on the right-hand side of Eq.~\eqref{eq:gonzalez_series}, corresponding to the first-order Euler method.
Gonz{\'a}lez \& Thompson~\cite{Gonzalez} then derived the starting formula
\begin{equation} \label{eq:gonzalez_init_cond}
    y_1 = y_0 + h y'_0 + \frac{h^2}{24} (7f_0 + 6f_1 - f_2) + O(h^5) \,,
\end{equation}
based on three-point forward finite differences to approximate the first and second-order derivatives in Eq.~\eqref{eq:gonzalez_series}.
Combining Eq.~\eqref{eq:gonzalez_init_cond} with the Numerov recurrence relation~\eqref{eq:numerov_recurrence} for the first iteration (i.e., $n=1$) results in the linear system
\begin{subequations} \label{eq:init_values_system}
\begin{align}
\Amat_2 \vb{y}_2 &= \vb{s}_2 \,, \\
\Amat_2 &=
    \begin{bmatrix}
 \KnXi{1}{(3)} & 1-\KnXi{2}{(\frac{1}{2})}\\
 -2 \KnXi{1}{-5} & \KnXi{2}{(1)}
\end{bmatrix} \,, \\
\vb{s}_2 &= 
\begin{bmatrix}
y_0  + h y'_0 + \frac{h^2}{24} \left( 7f_0 + 6 s_1 - s_2 \right) \\
-y_0  + \frac{h^2}{12} (s_2 + 10 s_1 + f_0)
\end{bmatrix}\,,
\end{align}
\end{subequations}
which can be simultaneously solved for $y_1$ and $y_2$; the algebraic solution is given by Eq.~(15) in Ref.~\cite{Gonzalez}. 
At this point, the first three grid points encoding the initial conditions, $\{y_0,y_1,y_2\}$, are determined. 
All other grid points, $y_{k\geqslant 3}$, are then obtained iteratively by the Numerov recurrence relation~\eqref{eq:numerov_recurrence} or likewise the matrix Numerov method outlined in Sec.~\ref{sec:matrix_numerov}.
This method was used in the \oneSzero and \threePzero channels in Ref.~\cite{Maldonado_thesis}.
However, one should be careful in higher partial-wave channels because the series expansion in Eq.~\eqref{eq:gonzalez_series} is only accurate up to $O(h^5)$.
Hence, it may not be accurate in higher partial-wave channels, especially those with $\ell \geqslant 4$.

\section{All-at-once Numerov for the \texorpdfstring{$T$}{T}-matrix element and other variations}
\label{app:other_emulators}

In this appendix, we present an alternative emulator for the complex-valued scattering $T$-matrix element.
The Numerov-based emulator equation reads
\begin{equation} \label{eq:app:lin_sys_T}
    \Amat(\paramVec) \vb{y}(\paramVec) = \vb{s}(\paramVec) \,,
\end{equation}
with $\yvec = \{ y_1, y_2,\ldots, y_N, T_\ell\}$ for a given initial value $y_0$. 
Note that $T_\ell$ is part of the solution vector $\yvec$. 
The solution vector $\yvec(\paramVec)$ is directly normalized according to the $T$-matrix asymptotic limit:
\begin{equation} \label{eq:asymLimitT}
    \phi_{\ell}(r;\paramVec) \sim  \frac{1}{p} \left[ F_\ell(pr) + T_{\ell}(\paramVec) \, H^+_\ell(pr) \right] \quad (r \to \infty)\,, 
\end{equation}
with the Hankel function $H^+_\ell(z) = G_\ell(z) + i F_\ell(z)$.
Motivated by the numerical linear algebra literature (e.g., see Refs.~\cite{Rees:2010,Stoll:2013}), we call this approach the ``all-at-once Numerov method'' because it solves the RSE subject to the asymptotic limit parametrization~\eqref{eq:asymLimitT} in one step.
From here on, we will omit the subscript indicating the angular momentum and instead use a subscript that specifies the sampling of these functions on the radial grid.
The corresponding $\Amat$ and $\vb{s}$ are
\begin{equation}
\begin{aligned}
\Amat &= 
\begin{bmatrix}
-2 \KnXi{1}{(-5)} & \KnXi{2}{(1)} \\
\KnXi{1}{(1)}     & -2 \KnXi{2}{(-5)} & \KnXi{3}{(1)} \\
                  & \ddots            & \ddots          & \ddots \\
                  &                   & \KnXi{N-2}{(1)} & -2 \KnXi{N-1}{(-5)} & \KnXi{N}{(1)} \\
                  &                   &                 & p                   & 0             & -H^+_{N-1} \\
                  &                   &                 &                     & p             & -H^+_N 
\end{bmatrix}
\,,\\
\vb{s} &= 
\begin{bmatrix}
- \KnXi{0}{(1)} y_0 + \frac{h^2}{12} (s_2 + 10 s_1 + s_0)\\
\frac{h^2}{12} (s_3 + 10 s_2 + s_1)\\
\frac{h^2}{12} (s_4 + 10 s_3 + s_2)\\
\vdots \\
\frac{h^2}{12} (s_N + 10 s_{N-1} + s_{N-2})\\
(1-\zeta p) F_{N-1} \\
(1-\zeta p) F_N  
\end{bmatrix} \,.
\end{aligned}
\end{equation}
As in Secs.~\ref{sec:fom} and~\ref{sec:rom}, $\zeta = 0$ ($\zeta = 1$) when solving the homogeneous (inhomogeneous) RSE.
We use again the short-hand notation for functions evaluated on the spacial grid, e.g., $H^+_{n} = H^+(p r_n)$.
The corresponding matrix in diagonal-ordered form reads:
\begin{equation}
    \bar{\Amat} = 
\begin{bmatrix}
* & \KnXi{2}{(1)}& \KnXi{3}{(1)}& \cdots & \KnXi{N}{(1)} & -H^+_{N-1}   \\
-2 \KnXi{1}{(-5)} & -2 \KnXi{2}{(-5)} & \cdots & -2 \KnXi{N-1}{(-5)} & 0& -H^+_N \\
\KnXi{1}{(1)} & \KnXi{2}{(1)} & \cdots & p & p & *
\end{bmatrix}\,.
\end{equation}
Since the $T$-matrix element is explicitly in $\yvec$, the ROM estimate $\yvec(\paramVec) =\Xmat \cvec(\paramVec)$ [see Eq.~\eqref{eq:rom_reduction}] simultaneously emulates both the wave function and the corresponding $T$-matrix elements, both of which are given by linear combinations of the snapshot solutions. 
While the linear combination of real $K$-matrices always leads to unitary $S$-matrices, a property of the Cayley transform, linear combinations of (complex-valued) $T$-matrices generally violate unitary.
However, the violation of unitary can be systematically reduced by improving the snapshot basis, e.g., using the developed greedy algorithm for basis selection, and thus may be irrelevant in practice.

Likewise, one can use any other scattering matrix element instead of the $T$-matrix element in the solution vector~\eqref{eq:app:lin_sys_T}; for example, the $K$-matrix element to avoid unitary violation.
However, in contrast to the $K$-matrix element, the absolute value of the $T$-matrix element is bounded for real-valued potentials: $|T| = K^2/(1+K^2) \leqslant 1$, thereby avoiding potential infinities in the emulator basis due to infinite $K$ matrices.
Hence, in this case, we formulate the Numerov recurrence relation, combined with the initial values for $y_0$ and $y_1$ discussed as the $(N+1)\times(N+1)$ linear system (as opposed to the $(N-1)\times(N-1)$ system discussed in Sec.~\ref{sec:matrix_numerov}) 
\begin{equation} 
    \Amat(\paramVec) \vb{y}(\paramVec) = \vb{s}(\paramVec) \,
\end{equation}
and solve it for $\vb{y} = \{ y_2, y_3,\ldots, y_N, a, b\}$.
Note that $\yvec(\paramVec)$ directly determines the coefficients $a$ and $b$ in the asymptotic limit:
\begin{equation} \label{eq:asymLimitT_ab}
    \phi_{\ell}(r;\paramVec) \sim  \frac{1}{p} \Bigl[ a_\ell(\paramVec) F_\ell(pr) + b_{\ell}(\paramVec) \, G_\ell(pr) \Bigr] \quad (r \to \infty)\,. 
\end{equation}
This is done by appending two rows and columns to the end of Eq.~\eqref{eq:matrix_numerov}, 
\begin{equation} 
\begin{aligned}
\Amat &= 
\begin{bmatrix}
\ddots        & \ddots    & \ddots \\
& \KnXi{N-2}{(1)} & -2 \KnXi{N-1}{(-5)} & \KnXi{N}{(1)} \\
&  & p & 0 & -F_{N-1} & -G_{N-1}\\
&  &  & p & -F_{N} & -G_{N}
\end{bmatrix} \,, \\ 
\vb{s} &= 
\begin{bmatrix}
\vdots \\
\frac{h^2}{12} (s_N + 10 s_{N-1} + s_{N-2})\\
(1 - \zeta p) F_{N-1}  \\
(1 - \zeta p) F_{N} 
\end{bmatrix} \,.
\end{aligned} 
\end{equation}
The corresponding matrix in diagonal-ordered form from Eq.~\eqref{eq:Abar} only has its last columns altered,
\begin{equation} 
 \bar{\Amat} = 
            \begin{bmatrix} 
\cdots & 0 & 0 & 0 & 0 & G_{N-1}\\
\cdots & \KnXi{N-2}{(1)}& \KnXi{N-1}{(1)} & \KnXi{N}{(1)} & F_{N-1} & G_N\\
\cdots & -2\KnXi{N-2}{(1)}& -2\KnXi{N-1}{(-5)} & 0 & F_N & *\\
\cdots & \KnXi{N-2}{(1)}& p & p & * & *\\
\end{bmatrix} \,,
\end{equation}
where zeros are explicitly included for clarity.
This formulation determines the $K$-matrix element via
\begin{equation}
    K(\paramVec) = \frac{b(\paramVec)}{\zeta p + a(\paramVec)} \equiv \frac{y_N(\paramVec)}{\zeta p + y_{N-1}(\paramVec)} \,.
\end{equation}
It is important to note that the coefficient vector $\cvec(\paramVec)$ is equivalent to the one obtained in Sec.~\ref{sec:matrix_numerov}. 
The extracted $K$-matrix elements are also equivalent if $\tau=2$; hence, the emulator in Sec.~\ref{sec:matrix_numerov} is a generalization of the one described in this appendix.

\section{Prestoring the scalar error in the offline stage}
\label{app:prestore_scalar_residual}

This appendix discusses prestoring tensors in the emulator's offline stage such that the magnitude of the scalar residual $\norm{\residual}^2$ can be efficiently reconstructed in the online stage.
Restating the definition of the residual in Eq.~\eqref{eq:residual_norm2},
\begin{align}
\norm{\romResidual}^2 &= (\svec - \Amat \ytilde)^\dagger (\svec - \Amat \ytilde) \, \notag\\
&= \ytilde^\dagger \Amat^\dagger \Amat \ytilde 
- 2\Re \left( \ytilde^\dagger \Amat^\dagger \svec \right)
+ \svec^\dagger \svec \,, \label{eq:norm_2_decom_init}
\end{align}
where $\Re(\bullet)$ denotes the real part of the complex-valued argument.
The cancellation in Eq.~\eqref{eq:norm_2_decom_init} can cause numerical artifacts due to finite-precision arithmetic, resulting in $\norm{\residual}^2 < 0$.
Therefore, one should check and strictly enforce that $\norm{\residual}^2 \geqslant 0$.

In the following, we rewrite the expression in Eq.~\eqref{eq:norm_2_decom_init} such that parts of it can be prestored.
To this end, let us recall the affine decompositions of the tensors:
\begin{align}
    \Amat_{ij} &= \sum_a \ATensor_{ija} \paramVec_a \,,\\
    \svec_{i} &=  \sum_a \sTensor_{ia} \paramVec_a \,.
\end{align}
We will omit the dependence on $\paramVec$ for brevity.

The first term on the right-hand side of Eq.~\eqref{eq:norm_2_decom_init} evaluates to $\ytilde^\dagger  \Amat^\dagger \Amat \ytilde = \cvec^\dagger \left[ \Xmat^\dagger (\Amat^\dagger \Amat) \Xmat \right] \cvec$, with
\begin{align}
\big(& \Xmat^\dagger  \Amat^\dagger \Amat \Xmat \big)_{uv}  \notag \\
&= \sum_{ab} \left( \sum_{ijk} (\Xmat_{iu}^* (\ATensor_{jia})^*  \ATensor_{jkb} \Xmat_{kv}\right)_{uvab} \paramVec_a^* \paramVec_b \,.
%
%
%
%
\end{align}
This rank-4 tensor has contains $(n_b^2 \times n_\theta^2)$ elements. 
The terms in parentheses can be prestored.

The second term can be expressed via:
\begin{align}
(\tilde{\svec}^\dagger \Amat \ytilde) 
&=  \sum_{abu} \left[ \sum_{ij} (\sTensor)_{ib}^* \ATensor_{ija}   \Xmat_{ju} \right]_{abu} \paramVec_a \paramVec_b^* \cvec_{u} \,.
%
%
%
%
\end{align}
%
The terms in brackets can be prestored. 

The third term evaluates to:
\begin{equation}
\svec^\dagger \svec 
=\paramVec^\dagger \left[(\sTensor)^\dagger \sTensor \right] \paramVec \,.
\end{equation}
Again, the term in the brackets can be restored.

\bibliographystyle{apsrev4-2}
\bibliography{bayesian_refs,more_refs,bib}

\end{document}